\documentclass[%
 reprint,
 amsmath,amssymb,
 aps,
prstper,
floatfix,
]{revtex4-2}

\usepackage{graphicx}% Include figure files
\usepackage{dcolumn}% Align table columns on decimal point
\usepackage{bm}% bold math
\begin{document}

\preprint{APS/123-QED}

\title{Course Deficit Model and the CLASP curriculum: Examining equity and graduation rates at two institutions}% Force line breaks with \\

\author{Cassandra A. Paul}
\affiliation{%
Physics \& Astronomy Department,
Science Education Program\\
San Jose State University %\textbackslash\textbackslash
}%
% \altaffiliation[Also at ]{Department of Physics & Astronomy Science Education Program, San Jose State University.}%Lines break automatically or can be forced with \\
\author{David J. Webb}%
\affiliation{%
 Department of Physics \& Astronomy,\\ University of California - Davis 
}%


%\affiliation{
%\affiliation{
%\affiliation{%


\date{\today}% It is always \today, today,

\begin{abstract}
We have previously described the reformed introductory physics course, Collaborative Learning
through Active Sense-Making in Physics (CLASP), for bioscience students at a large public research one university (Original University) and presented evidence that the course was more successful and more equitable than the course it replaced by several measures. Now we compare the original success of CLASP with an implementation at a second institution. We find that the original results hold at another institution despite some changes to the original curriculum and a somewhat different student population. We find that students who take CLASP are 1) less likely to drop, 2) less likely to fail, and 3) do as well in later coursework when compared to students who took the courses that CLASP replaced, even if that coursework is not similarly reformed. We find the above items to be independently true for historically marginalized students and remarkably, also find that 4) marginalized students who take CLASP are more likely to graduate from a STEM field. We use a course deficit model perspective to examine these results, and discuss some of the factors that may have contributed to this success. We argue that higher education has the tools needed to significantly increase equity, and improve student success and retention.

\end{abstract}

\maketitle

%\tableofcontents

\section{\label{sec:Intro}Introduction}

\subsection{\label{sec:Reform}Course Reform}

Retention in higher education has always been important, but it has taken on new significance now that enrollment in higher education is facing a pending enrollment cliff which has been further impacted by the effects of Covid-19 pandemic and other factors involving the changing demographics of the U.S \cite{schuette_navigating_nodate, copley_enrollment_2020, grawe_how_2021}.  One potential avenue of increasing retention is course reform, especially large introductory courses. In a large multi-institutional study, Hatfield, Brown \& Topaz \cite{Hatfield2022} find that when students struggle in one or more introductory STEM courses it significantly reduces their chances of graduation. Because there is overwhelming evidence that students perform better when active learning techniques are used in STEM classrooms \cite{Freeman2014b, kramer_establishing_2023}, reforming large undergraduate STEM courses can provide a powerful pathway towards increasing retention in higher education.

Course reform has been one of the major areas of study in Physics Education Research (PER) over the long history of our field. Often the results from the reformed courses are compared to the those of the more traditional courses that the reformed courses replaced whereas side-by-side comparisons \textbf{across different institutions} are less common \cite{Docktor2014}. We know curricular success depends partly on context, and that PER studies are primarily done at Research 1 (commonly referred to as R1) institutions and so are not representative of the general college population \cite{Kanim2020}. This suggests that PER is in need more replication studies at institutions that are significantly different from the institution of curricular origin.

Collaborative Learning through Active Sense-making in Physics (CLASP) is an active learning and model-based introductory physics reformed curriculum.  When compared to the traditional course, CLASP has been shown to be correlated with increased student GPA and MCAT scores, and resulted in high conceptual and attitudinal gains compared to national averages \cite{Potter2014Sixteen}.  CLASP has also been shown to increase the odds of course success for a variety of student demographics \cite{Paul2017} so that overall success improved under CLASP in an equitable way.  However, research on student success under this curriculum has been primarily limited to a single university.

In this paper we present a side-by-side analysis of CLASP implementations at two very different institutions, 20 years apart.  We show that there is a remarkable similarity across these schools in the ways that success in the course has become more equitable under CLASP.

\subsection{\label{sec:Active}Active Learning}
Over the past 30 years there have been many physics
course reform efforts (such as CLASP) featuring active learning via small-group peer-peer
discussions as a major course component. In practice
these peer-peer discussions often take place in a studio-type setting where students work together in small
groups on activities facilitated by the instructors \cite{Potter2014Sixteen, beichner_student-centered_2007,brewe_toward_2010,redish_nexusphysics_2014,etkina_investigative_2019}.  The PER community and the greater Discipline-Based Education Research (DBER) community have provided overwhelming evidence that active learning courses on average improve learning outcomes for students in these settings and beyond \cite{Freeman2014b, VonKorff2016, kramer_establishing_2023}.

However, research has also shown that there is a lot of variety in student performance on various metrics between different demographic populations in active learning environments \cite{etkina_lessons_1999, beichner_student-centered_2007, Cooper2016, Shafer2021, brewe_toward_2010, webb_attributing_2023}. For example, in their
2014 paper, Eddy and Hogan discuss how different populations interact and perform differently in active learning classrooms suggesting that there is a more nuanced answer to the question, ``Is active learning better?'' \cite{eddy_getting_2014}. They find that while all students performed better in an active learning environment, the gain in exam performance almost doubled for African American students and improved similarly for first generation college students. The African American students had different perceptions of the class than other groups, suggesting that the mechanism for increased performance may not be the same for all groups of students. Therefore, it’s important to consider for whom active learning is better, and by which metric.  In this paper we use several distinctly different metrics in order to develop a larger picture of equity issues and their changes under CLASP at these two institutions.

\section{\label{sec:Framing}Research Framing}

\subsection{\label{sec:CDM}Course Deficit Model}
In recent work \cite{webb_attributing_2023} we have suggested that demographic differences in course grades are best understood using a course deficit model \cite{Cotner2017}, rather than the more traditional student deficit model \cite{Valencia1997}.  In this paper we use an entirely different dataset and entirely different course changes in support of the same conclusion, that a particular course structure may contain hidden bias against some demographic groups so that a course deficit model is the best way to understand differences in results between different demographic groups.  Using a course deficit model means that we attribute the differences in performance between different students groups in a given course to the features of the course itself \cite{Cotner2017, webb_attributing_2023}.  The model is perhaps best understood in contrast to the student deficit model, which assumes that differences in course performance across different student demographic groups arise not from the course structure but from past inequities between different student demographic populations. Importantly, the course deficit model doesn't assume that students necessarily have equal backgrounds, it merely assumes that different student demographic groups are equally capable of succeeding in a given course, and that any gaps in performance are instead the result of different classroom practices systematically favoring some populations over others.

\subsection{\label{sec:Equity}Equity Models}
Since equity can mean different things to different people and in different contexts, we will be careful with our meaning in this paper.  In this work we apply two models of equity; Equity of Parity and Equity of Individuality \cite{Rodriguez2012}.  Equity of Parity works particularly well with a course deficit model, because in order for equity of parity to be achieved, all demographic groups must achieve the same outcome. Importantly, when Equity of Parity is used with a course deficit model, it doesn't ignore past inequities suffered by different demographic groups, it merely indicates that these inequities are inconsequential to success in the course. In this paper we consider Equity of Parity achieved when the average success of individual demographic groups are  statistically indistinguishable from the average of all other groups \cite{Rodriguez2012, Gutierrez2012}.  We also use the model of Equity of Individuality. This model of equity compares each groups' success to the prior success of that group instead of comparing to a specific comparison group \cite{Rodriguez2012}.  We will be careful to examine the impact of our interventions on many different student demographics in order to ensure that the intervention doesn't inadvertently disadvantage some groups even while increasing overall gains. 

While it is customary in educational research to group all marginalized groups together into one single group for statistical purposes, this practice can hide important differences between demographic groups. Following the work of Shafer et. al. \cite{Shafer2021} as well as our own \cite{Paul2022}, we recognize that different metrics yield different results among different student demographic groups. In our work in this paper, we are careful to examine the impacts of our intervention on many different student demographic groups. Because we want to ensure simplicity in our message, we have chosen to present marginalized student groups together when the outcome is similar for all groups. However, we include the comparisons by detailed student demographics in the appendix for completeness.

\section{\label{sec:Questions}Research Questions}
In this paper we examine several years of transcript
data for students across different demographic groups in order
to better understand how the switch from a traditional
course to the active-learning based Collaborative
Learning through active Sense-making in Physics
(CLASP) curriculum \cite{Potter2014Sixteen, Paul2017, Harrer2019a} has impacted different groups of
students at two different institutions. We have previously shown that CLASP produces better learning gains on average than the traditional course it replaced at a large public R1 institution \cite{Potter2014Sixteen} and, in a conference proceeding, we also showed that the impact of the CLASP course at the original institution was positive across all ethnic demographic groups we examined \cite{Paul2017}.  Since CLASP was implemented several years ago at a second institution, this study is primarily a replication study of that conference proceeding. We show side-by-side results from both institutions.

As a part of a larger project, in this study we consider equity across student ethnicity by examining course grades and
pathways towards graduation.  We examine the success
of students grouped by self-identified ethnicity (as
defined and collected by each institution), comparing success
in CLASP with that of the traditional course it replaced.  The practice of examining improvements within specific student
populations, without comparing their success to that of
some normative group, is referred to as applying a
model of Equity of Individuality \cite{Rodriguez2012}.  We say more
about how this model is used in our results and in our discussion.

We investigate the following research questions:

\begin{enumerate}
  \item How do students of different demographic groups perform in the CLASP series as compared to the traditional course it replaced at each institution?
  \item What evidence exists that CLASP students of different demographics are prepared for later coursework?
  
  \item Are CLASP students of some demographics more likely to graduate than those who took the course that CLASP replaced? 
\end{enumerate}

\section{\label{sec:Setting}Course Setting}
As described in our previous work \cite{Potter2014Sixteen}, the CLASP
curriculum at the Original University (OU), was implemented in an introductory physics course required
for many STEM degrees, including the biosciences. CLASP features an explicit focus on understanding models (including words, graphs, equations, etc.) and
using these models in qualitative and quantitative analysis of real-world physical, chemical, and biological situations \cite{Potter2014Sixteen, Harrer2019a}.  The active learning elements of the course are carried out in studio-type discussion/laboratory (DL) sections of 25-35 students. These DL sections
include small group activities where students work together to understand a model and practice applying it before engaging in whole-class discussions. At Original University, CLASP is a three-quarter (one year) series, meaning there are
CLASP Physics A, B and C courses. 

At both Next and Original Universities, the students meet for 140 minutes twice a week to work through activity cycles of small group work, and whole class discussions. Lecture (during the time of the study on both campuses) was limited to one 80 minute session per week. DLs are taught by either a faculty member or a graduate teaching assistant. At both locations, the lecture (typically taught by a faculty member) reviews or previews the the topics discussed in the DL, and does not introduce material not covered in the DLs. The DLs are considered the main driver of the curriculum. Exams and quizzes, which were conducted during the lecture section, determine most of the students' grades. 

There are a few differences between the implementations of CLASP at the two universities. At Next University, the introductory series is also a year-long course but, since the Next University is on semesters, it is a two-semester series. An additional difference between the two universities is that at Next University only the A course is based on CLASP, the B course is a traditional course for the years in this study.

There are also differences in the order of the material in the two courses. Both universities (non-traditionally) begin their curricula with fundamentals concerning the conservation of energy including the three-phase model of matter, then introducing phase and thermal energies, then move into mechanical models of kinetic and potential energy. However, while the Original University covers particle models of bond and thermal energy, and then moves into thermodynamics and the ideal gas model during the first quarter, Next University skips particle models and much of the thermodynamics, and instead moves to more traditional introductory physics concepts like conservation of momentum, Newton's laws, circular motion and kinematics. (Original University covers these topics in their B section of the course using the same materials.) These curricular ordering decisions were made so that the changes made to Next University's course did not change the content covered in either the A or the B section course. 

Next University's discussion/laboratory (DL) sections are also generally smaller, with the maximum number of seats being 25 instead of Original University's maximum of 35. Finally, beginning in 2016, Next University also had undergraduate Learning Assistants in many DL and lecture sections.

\section{\label{sec:Methods}Methods}

\subsection{\label{sec:Grades}Grades and Retention}

Some studies have shown that grades are not necessarily good indicators of learning because students have an
ability to solve physics problems without a good
understanding of the physics content \cite{Kim2002Students,crouch_peer_2001}, therefore
student pass-rates may not be a good indicator of
learning goals. However, grades still play an important
role in the academic process. For example, because
grades represent an official form of evaluation, good
grades can encourage students, and poor grades can
deter them [13]. Furthermore, pass-rates are a very real
indicator of college success, as course failures can mean
students need to repeat coursework, drop out of their
major, and/or leave college entirely. 

While we will be using the absolute grade a student gets as one measure of success we will not be using past grades as controls for any model that makes comparisons across demographics.  It is common in quantitative PER for researchers to choose to control for students' prior preparation or grade point average.  Because we are using a course deficit model, we do not intend to do this.  This is for two reasons.  First, as described earlier, the Course Deficit Model ignores incoming differences and attributes any differences in course performance between student demographic groups to course structures.  We expect to see differences in student preparation resulting from the structural racism evident in society.  Thus, controlling for preparation could potentially hide the equity gaps we are trying to address \cite{webb_attributing_2023,castillo_wendy_how_2023}.  Second, the grades that students receive in a given course have very real implications for students.  Take this illustrative example: if one demographic group receives an average of D in a course, and another demographic group receives and average of an B, but when we control for preparation then both groups perform equitably (as they both have the same expected growth), this hides the very different experiences and outcomes that each demographic group has in that classroom, and any resulting consequences of those differences (i.e. needing to re-take the course, getting behind, and even dropping the major). Fortunately the Course Deficit Model provides the best explanation for differences in the success of different demographic groups so these problems with a Student Deficit Model need not arise.

\subsection{\label{sec:OU}Original University Context}
Original University began offering trials of
CLASP in 1995, and completely replaced the traditional lecture/lab course by academic year 96-97. This switchover creates the conditions of a natural experiment that allowed us to compare student successes
in the traditional lecture/lab courses with successes in an active-learning studio-type courses (CLASP).  For all analyses we will use the same set of students at Original University that we used for our earlier paper \cite{Paul2017}, comparing those who took CLASP to those who took the traditional course CLASP replaced.  The NonCLASP classes we include were offered from the Fall quarter of 1993 through the Summer of 1996 and the CLASP courses included were offered from Winter quarter of 1995 through Summer of 2000.

Ethnicity data are self-reported by all students from pre-determined ethnicity categories created and named by Original University. Students were only allowed to select one option but they could choose not to select any.  The student population
represented in the data is 84\% US citizens and 16\% permanent residents.  Groups historically marginalized in physics \cite{APSdiversity} (we will use the acronym HM for this demographic grouping) in this study include African Americans, Latina(o), Chicana(o) or
Mexican American, Hispanic-other, Native American, and Pacific Islander students. The population fitting the HM classification is about 89\% US citizens and 11\% permanent residents.

The regular terms offered from Fall 1993-Summer 2000 included 8235 students (of which we have ethnicity information for 7896). Table \ref{tab1a} gives a short demographic breakdown of these students over these years.
\begin{table}[htb]
\caption{Some demographics of our dataset from OU.  The fractions of Male students, HM students, and STEM majors in each set of either NonCLASP or CLASP courses is shown.}
\label{tab1a}
\begin{ruledtabular}
\begin{tabular}{c c c c c }
\textbf{Group} & \textbf{N} & \textbf{Male} & \textbf{HM} & \textbf{STEM} \\ 
 \hline
 NonCLASP & 3386 & 0.44 & 0.12 & 0.90 \\
CLASP & 4849 & 0.38 & 0.13 & 0.93 \\
\end{tabular}
\end{ruledtabular}
\end{table}

\subsection{\label{sec:NU}Next University Context}

Trial runs of the CLASP courses at the next university (NU), following on the work at the original university (OU), were begun in 2013 and continued through 2014, running at the same time as the NonCLASP courses that would be replaced.  Only the first semester (Physics A) of the two-semester series was changed to the CLASP format. To include enough students for analyses we use these two years that had both versions being offered and also add in the attached three years, 2010-2012, of no CLASP courses and the three years, 2015-2017, of only CLASP courses.

Summer classes didn’t ever change to CLASP format but they make up only 3\% of the total number of students which is not large enough to be a useful comparison group by themselves. Because these summer classes are also likely to include the highest percentages of non-standard students (students from other universities, students trying to get out of academic probation, etc.), we will leave them out of all of our analyses and just compare the regular term CLASP with regular term NonCLASP students.  This will change the numbers slightly but not any of our conclusions.

The regular terms offered from Spring 2010-Fall 2017 included 3219 students (of which we have ethnicity information for 2943). Table \ref{tab1} gives a short demographic breakdown of these students over these years.

\begin{table}[htb]
\caption{Some demographics of our dataset from NU.  The fractions of Male students, HM students, and STEM majors in each set of either NonCLASP or CLASP courses is shown.}
\label{tab1}
\begin{ruledtabular}
\begin{tabular}{c c c c c }
\textbf{Group} & \textbf{N} & \textbf{Male} & \textbf{HM} & \textbf{STEM} \\ 
 \hline
 NonCLASP & 1927 & 0.52 & 0.24 & 0.68 \\
CLASP & 1292 & 0.55 & 0.28 & 0.68 \\
\end{tabular}
\end{ruledtabular}
\end{table}

\section{\label{sec:Drops}Dropping the course}

\subsection{\label{DropGraphs}Graphical Examination of Drop-rates}

\begin{figure*}[htb]
\includegraphics[trim=1.1cm 1.5cm 0.8cm 1.3cm, clip=true,width=\linewidth]{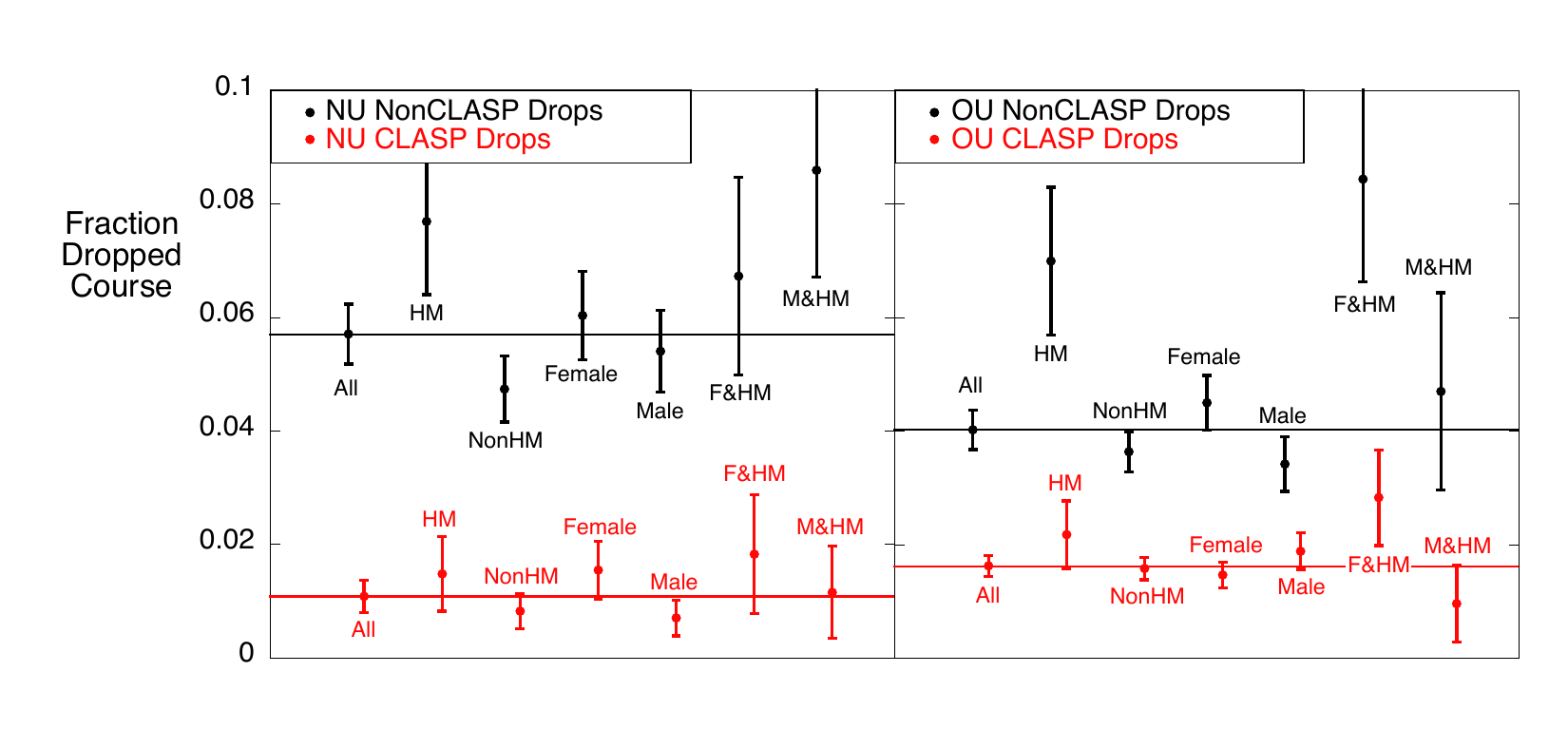}
\caption{The fraction of drops in NU CLASP and NonCLASP classes is shown on the left.  Similar data for OU is shown on the right.  The error bars are standard errors.  In each case several different way of separating the students into demographic groups are shown.  Specifically, two intersections of the other groups, are also shown.  For example, F\&HM refers to the group of students who identified as female and also as members of racial/ethnic groups historically marginalized in physics.}
\label{fig:Drops}
\end{figure*}

To succeed in a class, the first thing a student enrolled in the class must do is to not withdraw from the class.  So, first we'll examine the fractions of students who dropped the two types of classes, how these fractions can depend on some demographic variables, and compare the CLASP/NonCLASP results for the two very different universities separated by about 20 years in time.  Fig.\ref{fig:Drops} shows these drop fractions with a set of CLASP/NonCLASP comparisons for Next University on the left and for Original University on the right.  To make the comparisons of equity across different demographic groups easier to do by eye, we have drawn a horizontal line through the overall drop fraction for CLASP course in red and NonCLASP courses in black.  One sees that the CLASP courses at both universities a) have  considerably smaller fractions of students who choose to drop the course and that b) these drop fractions have less of a dependence on demographic group than the NonCLASP courses. We examine these two findings statistically in the following section. It is also rather striking that these two results show up in a similar way for two very different universities 20 years apart. (For further graphical breakdown of demographic groups by gender and race/ethnicity, see Fig. \ref{fig:EthGenOUDrops} \& Fig.
\ref{fig:EthGenNUDrops} in Appendix \ref{sec:DropsAllEthnicities}.)

\subsection{\label{DropStats}Statistical Examination of Drop-rates}

\subsubsection{\label{DropsBetween}Drop-rate changes between NonCLASP \& CLASP courses}

For all students, the odds of dropping a NonCLASP course are significantly higher than the CLASP course at both universities. Table \ref{tab2a} shows these results along with the effect size of about 19 fewer CLASP students dropping per year and about 30 fewer CLASP students dropping per year at Original University. 

In addition to finding fewer students dropping over all, we can use an Equity of Individuality model of equity, to examine the differences between CLASP and NonCLASP courses for \textbf{each} demographic group.  To quantify the CLASP/NonCLASP differences we see in Fig. \ref{fig:Drops} we use logistic regression to estimate the ratio of the odds of a group member dropping a CLASP course to the odds of a member of the same group dropping a NonCLASP course.  Eq. \ref{eqn:Odds} defines the odds of an event and our model is shown in Eq. \ref{eqn:DropOdds}.  We use this model to fit our data separately for each demographic group:
\begin{equation}
\text{Odds of Dropping} = \frac{\text{Drop Probability}}{1-\text{Drop Probability}} 
\label{eqn:Odds}
\end{equation}

\begin{equation}
Log(DropOdds) = b_0 + b_{CLASP}CLASP 
\label{eqn:DropOdds}
\end{equation}
where CLASP = 1 if the student was enrolled in a CLASP class and 0 if they were in a NonCLASP class.  Eq. \ref{eqn:Ratio} is used to convert a coefficient from the fit into an odds ratio.
\begin{equation}
\begin{split}
\text{Odds Ratio} & = e^{b_{CLASP}} \\ & = \frac{\text{CLASP DropOdds}}{\text{NonCLASP DropOdds}}
\end{split}
\label{eqn:Ratio}
\end{equation}
Table \ref{tab2a} shows the odds ratios and standard errors for each fit of the model to measure the difference between CLASP classes and NonCLASP classes for each demographic group shown in Fig. \ref{fig:Drops}.  In addition we have given the estimated net number of students per year (along with the standard error) who would have dropped if they were enrolled in a NonCLASP course but not if they were enrolled in a CLASP course.  If CLASP and NonCLASP course produce equal odds of dropping the course we would find an odds ratio = 1.  As expected, all of these odds ratios are less than one with the smallest odds ratios across both universities for male students from historically marginalized groups showing that at NU they had eight times the odds of dropping NonCLASP as of dropping CLASP and at OU five times the odds of dropping NonCLASP compared to dropping CLASP.  Again, these two specific similar results are from two different universities 20 years apart.

\begin{table*}[htb]
\caption{\textbf{Class-type comparisons of course dropping, for specific groups}.  For each university and for each demographic group in Fig. \ref{fig:Drops} we list the number of students in the measurement, the odds ratio for dropping = (odds of dropping a CLASP course) / (odds of dropping a NonCLASP course), the standard error for those odds ratios, the P-value for those odds ratios, and the estimated net number of students \textbf{per year} from that demographic group who would have dropped from the NonCLASP course but would benefit from the CLASP course by not dropping their class.  The standard error on this number of students is in parentheses.}
\label{tab2a}
\begin{ruledtabular}
\begin{tabular}{c | c c c c | c c c c}
\textbf{Group} & \multicolumn{4}{c |}{\textbf{NU}} & \multicolumn{4}{c}{\textbf{OU}} \\
 &  &  &  & \textbf{Extra NonCLASP} &  &  &  & \textbf{Extra NonCLASP} \\ 
 & \textbf{\textit{N}} & \textbf{Odds Ratio (SE)} & \textbf{P-value} & \textbf{yearly drops (SE)} & \textbf{\textit{N}} & \textbf{Odds Ratio (SE)} & \textbf{P-value} & \textbf{yearly drops (SE)} \\ 
 \hline
\textbf{All} & 3219 & 0.181 (0.052) & $<10^{-3}$ & 18.6 (2.4) & 8235 & 0.404 (0.056) & $<10^{-3}$ & 29.5 (4.7) \\
\textbf{HM} & 766 & 0.181 (0.088) & $<10^{-3}$ & 5.9 (1.4) & 990 & 0.30 (0.10) & $<10^{-3}$ & 6.8 (2.0) \\
\textbf{NonHM} & 2177 & 0.167 (0.067) & $<10^{-3}$ & 10.7 (1.8) & 6906 & 0.425 (0.069) & $<10^{-3}$ & 21.0 (4.0)\\
\textbf{Female} & 1510 & 0.245 (0.089) & $<10^{-3}$ & 8.5 (1.8) & 4867 & 0.315 (0.059) & $<10^{-3}$ & 21.6 (3.7)\\
\textbf{Male} & 1709 & 0.124 (0.058) & $<10^{-3}$ & 10.0 (1.7) & 3304 & 0.569 (0.125) & $<10^{-3}$ & 6.9 (2.7)\\
\textbf{F\&HM} & 372 & 0.26 (0.17) & 0.036 & 2.3 (0.9) & 628 & 0.32 (0.12) & 0.003 & 5.0 (1.8)\\
\textbf{M\&HM} & 394 & 0.124 (0.093) & 0.005 & 3.7 (1.0) & 362 & 0.20 (0.16) & $<10^{-3}$ & 1.9 (1.8)\\
\end{tabular}
\end{ruledtabular}
\end{table*}

\subsubsection{\label{DropsWithin}Equity of Drop-Rates Within Class-types}

Next, we can think of equity within each particular class type using an Equity of Parity model of equity.  We will be using the idea of Guti{\'{e}}rrez \cite{Gutierrez2012}, that demographic equity means that a student's success shouldn’t be predictable from their demographic characteristics.  First we have to say that inequity is easier to quantify than equity.  For instance, in NonCLASP courses at NU the data show us that 7.7\% of students from historically marginalized groups dropped the course but only 4.7\% of their peers dropped the course.  That difference amounts to about 3  \textbf{extra} marginalized students, on average, dropping the course every year.  If we assume that the NonCLASP courses were fully equitable and use a $\chi^2$ test (the details of these $\chi^2$ tests can be found in Appendix \ref{sec:DropPassGradEquity}) to measure how likely it is that the extra drops by marginalized students are just a statistical fluctuation then we find that the there is only a 2\% chance (a P-value of 0.02) of a statistical fluctuation this large for a course that meets Equity of Parity.  Most researchers would use values of P smaller than about 0.05 as evidence that this NonCLASP course \textbf{is not equitable} \footnote{Within the Course Deficit model the cause of the inequity is course structure.  Hence the conclusion that the course is inequitable}.  On the other hand, in the NU CLASP courses there was less than one \textbf{extra} marginalized student per year dropping the course and a $\chi^2$ test gives us P = 0.31 so in this case, on the assumption that CLASP courses are equitable, we have a 31\% chance that this result is a statistical fluctuation.  For this P-value we would say that the data don't provide enough evidence to conclude that the course is inequitable (which is not quite the same as concluding that the course is fully equitable).  In Appendix \ref{sec:DropPassGradEquity} we use $\chi^2$ tests for each of the demographic groups in Fig. \ref{fig:Drops} and their peers to quantify what we can already see fairly clearly from the figure itself.  Specifically, the figure shows that the groups in the CLASP courses have drop fractions all roughly within a standard error of the overall average.  Conversely, in the NonCLASP courses some groups have drop rates considerably farther from average.  Even though we can't prove the CLASP courses are equitable, we do use these results to suggest that CLASP is \textbf{more} equitable than NonCLASP, at least in terms of who drops the course.

\section{\label{sec:PassRates}Passing the course}

The second step in succeeding in the first course of these intro-physics series (after not dropping the course) is to receive a grade of C- or better so that the course needn't be repeated. So, the next comparisons we make between these courses are finding the fractions of students who succeeded by receiving C- or higher grades in the first course in the sequence.  We also examine equity for this type of success.

\subsection{Graphical Examination of Passing}

\begin{figure*} [htb]
\includegraphics[trim=0.9cm 1.3cm 1.0cm 1.3cm, clip=true,width=\linewidth]{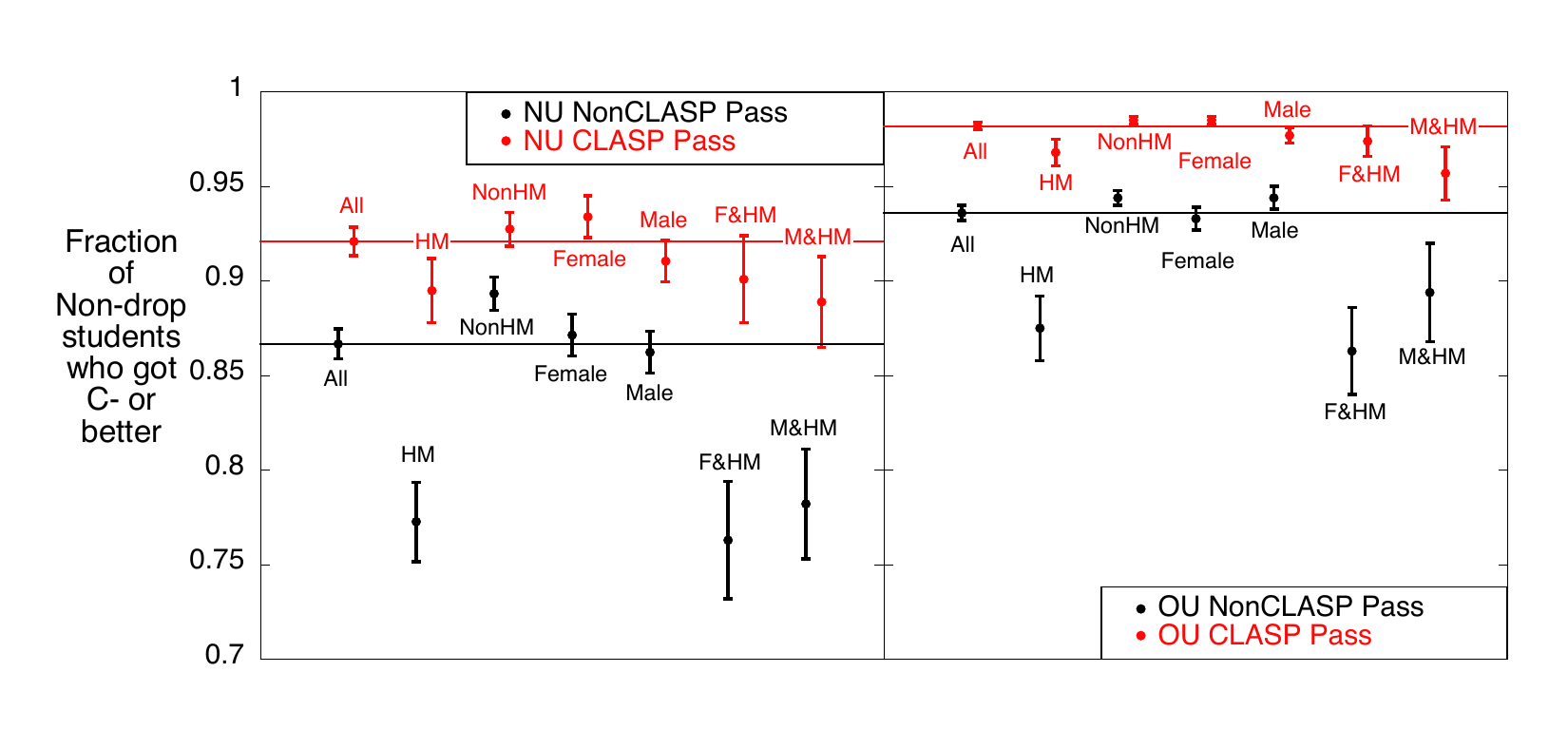}
\caption{The students who didn't drop the course received grades.  This plot shows the fraction of those students who received a grade of C- or better (in some important ways this amounts to passing the course).  These grade fractions are shown for NU CLASP and NonCLASP classes on the left and for OU on the right.  The error bars are standard errors.  Again, several different ways of separating the students into demographic groups are shown.}
\label{fig:Pass}
\end{figure*}

Fig. \ref{fig:Pass} shows the fraction of students that received a grade higher than D+ and therefore passed the course. The CLASP courses at both universities saw the highest level of success by this measure, indicated by the red lines being higher than the black lines in both cases. In addition to having these higher success levels overall, the CLASP courses appear to be the more equitable than nonCLASP in the sense of the Equity of Parity model because the averages for each demographic group appear much closer to the average line for the students who took CLASP courses averages than they do for the students who took nonCLASP courses. Similarly, we can use an equity of individuality model and see that for each demographic group shown, the CLASP course shows an improved pass-rate, thus satisfying Equity of Individuality. (For further graphical breakdown of demographic groups by gender and race/ethnicity, see Fig. \ref{fig:EthGenOUPass} \& Fig.
\ref{fig:EthGenNUPass} in Appendix \ref{sec:PassAllEthnicities}.)

\subsection{Statistical Examination of Passing the Course}

\subsubsection{Pass-rate changes between NonCLASP and CLASP courses}

In Table \ref{tab3a}, we show that significantly more students are passing overall in the CLASP courses. At Next University we see an effect size of about 21 more students per year passing, and at Original University we see about 53 more students passing per year. The impact is similar given that Original University enrolls more students per year.

We can use the Equity of Individuality model to examine the impact on each demographic group.  We use Eq. \ref{eqn:DropOdds} but with DropOdds replaced with PassOdds and fit each demographic group in Fig. \ref{fig:Pass} including all of the students who did not drop and so received a grade. The results for odds ratio = (odds of passing CLASP with grade of C- or better)/(odds of passing NonCLASP with grade of C- or better) and standard error are given in Table \ref{tab3a}.  This table also gives the estimated net number of students per year (along with the standard error) who would have failed a NonCLASP course but passed a CLASP course.  As expected from the figure all of these odds ratios are larger than one. The largest odds ratio across both universities is for female students from historically marginalized groups who had about 3 times higher odds of passing CLASP at NU and about 6 times higher odds of passing CLASP at OU.  Again, similar results at very different universities.

\subsubsection{Equity of Pass-rates Within Class-types}

We have already noted the difficulty in using quantitative methods to prove Equity of Parity.  For this reason, our quantitative analyses regarding Equity of Parity will be presented in Appendix \ref{sec:DropPassGradEquity}.  However, we also note that this type of equity is already visible to the eye in Fig. \ref{fig:Pass} where we see that only for the CLASP courses does each demographic group have pass rates near the overall average so, although we may not argue that CLASP is fully equitable, it does appear that CLASP courses are more equitable than NonCLASP.

\begin{table*}[htb]
\caption{\textbf{Class-type comparisons of passing the course, for specific groups}.  For each university and for each demographic group in Fig. \ref{fig:Pass} we list the number of students in the measurement, the odds ratio for passing the course = (odds of passing a CLASP course) / (odds of passing a NonCLASP course), the standard error for those odds ratios, the P-value for those odds ratios, and the estimated net number of students \textbf{per year} from that demographic group who would not have passed the NonCLASP course but would benefit from the CLASP course because they passed their class.  The standard error on this number of students is in parentheses.}
\label{tab3a}
\begin{ruledtabular}
\begin{tabular}{c | c c c c | c c c c}
\textbf{Group} & \multicolumn{4}{c |}{\textbf{NU}} & \multicolumn{4}{c}{\textbf{OU}} \\
 &  &  &  & \textbf{Extra CLASP} &  &  &  & \textbf{Extra CLASP} \\ 
 & \textbf{\textit{N}} & \textbf{Odds Ratio (SE)} & \textbf{P-value} & \textbf{passes/year (SE)} & \textbf{\textit{N}} & \textbf{Odds Ratio (SE)} & \textbf{P-value} & \textbf{passes/year (SE)} \\ 
 \hline
\textbf{All} & 3095 & 1.79 (0.22) & $<10^{-3}$ & 21.1 (4.2) & 8004 & 3.75 (0.49) & $<10^{-3}$ & 52.6 (5.4) \\
\textbf{HM} & 728 & 2.50 (0.54) & $<10^{-3}$ & 11.1 (2.5) & 950 & 4.3 (1.2) & $<10^{-3}$ & 12.5 (2.6) \\
\textbf{NonHM} & 2107 & 1.53 (0.25) & $<10^{-3}$ & 9.0 (3.3) & 6733 & 3.84 (0.60) & $<10^{-3}$ & 39.0 (4.6)\\
\textbf{Female} & 1445 & 2.07 (0.41) & $<10^{-3}$ & 11.2 (2.8) & 4736 & 4.79 (0.86) & $<10^{-3}$ & 35.6 (4.3)\\
\textbf{Male} & 1650 & 1.63 (0.26) & 0.003 & 9.9 (3.2) & 3216 & 2.54 (0.50) & $<10^{-3}$ & 15.2 (3.2)\\
\textbf{F\&HM} & 355 & 2.82 (0.88) & 0.001 & 6.1 (1.7) & 597 & 5.9 (2.2) & $<10^{-3}$ & 9.5 (2.1)\\
\textbf{M\&HM} & 373 & 2.23 (0.66) & 0.007 & 5.0 (1.8) & 353 & 2.6 (1.1) & $<10^{-3}$ & 3.2 (1.5)\\
\end{tabular}
\end{ruledtabular}
\end{table*}

\section{\label{sec:Prepared?}CLASP students prepared for later study}
The algebra-based intro-physics series at NU consists of two semesters but only the first semester, PhysicsA, was changed to the CLASP format. The notably low drop rates, higher pass rates, and CLASP curriculum itself (which covers somewhat different content, and emphasizes different skills than the traditional course it replaced) leaves open the possibility that students may be leaving the CLASP course insufficiently prepared for the following PhysicsB non-CLASP course and, indeed, for the rest of their work in their fields. We can quantify the large demographic effects on PhysicsB that may be due to CLASP-PhysicsA's low drop and fail rates.  The CLASP PhysicsA courses fed 68\% $\pm$ 2\% of their nonHM students into PhysicsB and 67\% $\pm$ 3\% of their HM students.  In contrast the NonCLASP PhysicsA courses that were supplanted sent 67\% $\pm$ 1\% of their nonHM students but only 55\% $\pm$ 2\% of their HM students. So the inequities of the NonCLASP courses and the equity of the CLASP courses are obvious in this measure. It is also clear that the CLASP PhysicsA courses changed the demographics of the PhysicsB courses.  First, we'll look at average grades in PhysicsB at NU and then at the upper division GPAs of these students in order to find out if there is a noticeable effect of sending some 22\% more HM students into PhysicsB and beyond.

\begin{figure} [htb]
\includegraphics[trim=3.5cm 3.5cm 5.5cm 3.5cm, clip=true,width=\linewidth]{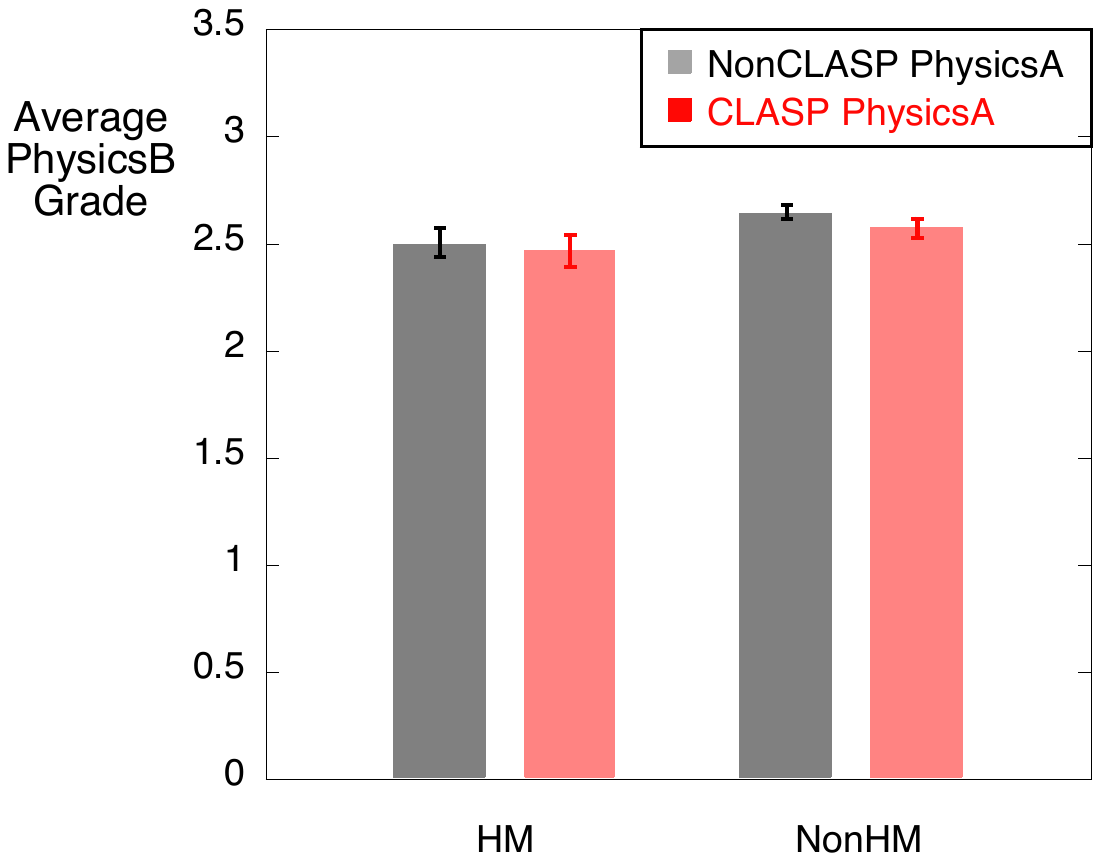}
\caption{The CLASP versions of the PhysicsA course sent over 20\% more HM students into PhysicsB than the NonCLASP versions of PhysicsA but the entire set of CLASP students appeared to be as prepared as the NonCLASP students to succeed at PhysicsB (which was never changed to the CLASP format).}
\label{fig:2BGrades}
\end{figure}

Because of the separation in time, between the NonCLASP PhysicsA courses, 2010-2014, and the CLASP PhysicsA courses, 2014-2017, these students took different PhysicsB courses.  So, first we wanted to test whether there were any noticeable grading changes in the PhysicsB courses from the period 2010-2014 to the period 2014-2017.  Fortunately, these courses all include a group of students who have never enrolled in any PhysicsA course, because they took a course equivalent to PhysicsA somewhere else.  This group of students can be used, as a kind of standard, to compare the PhysicsB courses across time.  Over 2010-2014 these students had an average PhysicsB grade of 2.71 $\pm$ 0.06 and over 2014-2017 they had an average grade of 2.68 $\pm$ 0.06 so it appears that there is no measurable change in the grading over these years.  The NonCLASP PhysicsA NonHM students had 2010-2014 an average PhysicsB grade of 2.65 $\pm$ 0.03 and the HM students had an average grade of 2.51 $\pm$ 0.07.  For the CLASP PhysicsA students over 2014-2017 those numbers were 2.57 $\pm$ 0.04 for NonHM students and 2.47 $\pm$ 0.07 for HM students.  Figure \ref{fig:2BGrades} shows these results.  We suggest that these small effects (Cohen's d of $0.08\pm0.06$ for NonHM and $0.03\pm0.17$ for HM) between the CLASP and NonCLASP students in PhysicsB, even if they are not simply statistical fluctuations, show that the CLASP program is not only more equitable but also provides enough background for the students to succeed in their next physics course.
One point we want to emphasize here is that despite the fact that CLASP A at Next University sent 22\% more HM students to Physics B, this did not result in a statistically different outcome grade-wise for HM students.

Next we examine upper division GPA's for these students to compare CLASP PhysicsA students with NonCLASP PhysicsA students.  We want to compare students taking roughly the same upper-division major courses so we will narrow down the group to students who majored in the 19 subjects that make up almost 90\% of the majors taking the course.  The graduation years 2017-2018 had about the same number of CLASP and NonCLASP students graduating so we will examine the GPAs for the final 2.5 years of these students' careers and take these GPAs as a proxy for their upper division major GPA.  We only include students who had at least 4 terms of study after taking PhysicsA, who had at least 43 units in that time, who didn't transfer more than 20\% of their units during this time, and we remove any PhysicsA grade points and units from the GPA.  Our resulting sample size is N = 487.  We find that the NonCLASP students graduating in those two years had an average upper-division GPA of 3.02 $\pm$ 0.03 and the CLASP students had an average GPA of 3.16 $\pm$ 0.03.  It might seem like the CLASP students thus did better in their later courses but, unfortunately, there is a confounding variable that seems to produce this apparent success.  A look at the data shows us clearly that students graduating in these two years had higher GPAs if they took PhysicsA closer to graduation.  This amounts to some sort of time-to-degree measure and turns out to hold for students from both the CLASP and NonCLASP courses.  If we control for this time-to-degree-like variable then the difference in upper-division GPA between the CLASP and NonCLASP courses becomes 0.022 $\pm$ 0.043 so the GPA effect is removed and we conclude that the two groups are roughly equally prepared for upper-division work.  This result does not change if we also control for student academic ability by controlling for a student's GPA upon entering PhysicsA.

So, for both later physics work as well as later work in their major it appears that the students who passed through CLASP courses were prepared as well as the students who passed through the NonCLASP courses.  Of course one might already think that one semester will not measurably change the academic trajectory of a group of students, however, we will see that by a different measure, STEM graduation rates, even one semester may have an effect.

\section{\label{sec:GradSTEM}Staying and graduating in STEM}

Previous work has suggested that grades in introductory STEM courses may well affect graduation rates, particularly for students from underrepresented groups \cite{Hatfield2022}. The years included in our datasets are all long enough ago that we can expect to measure graduation rates for the students in our dataset so that we can compare STEM graduation rates for CLASP and NonCLASP courses. Again, we note that at the Original University an entire year of courses was changed to CLASP but that at the Next University only the first semester of the algebra-based introductory physics series was changed to the CLASP format.

\begin{figure*} [htb]
\includegraphics[trim=1.2cm 1.6cm 1.4cm 1.3cm, clip=true,width=\linewidth]{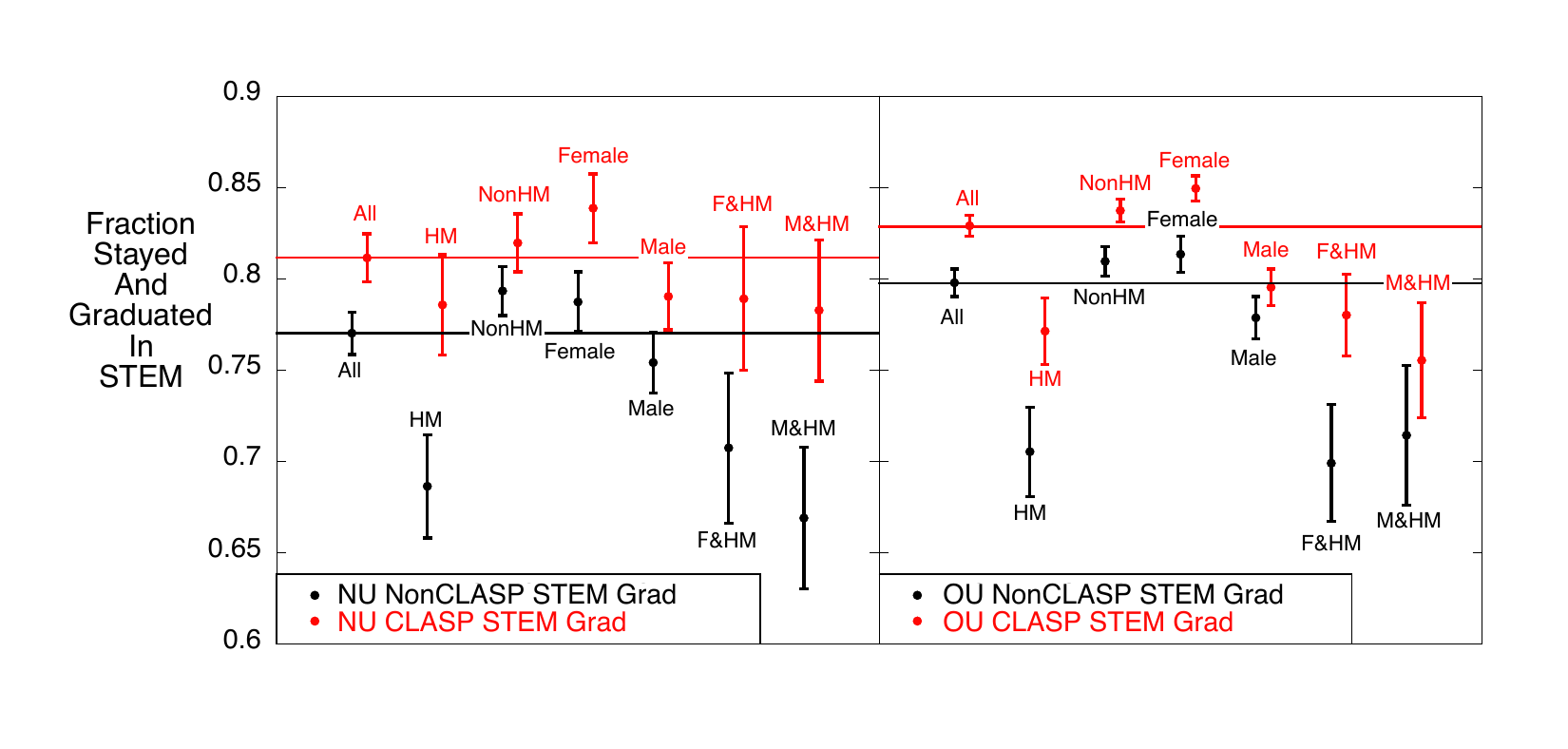}
\caption{Students who were STEM majors when they enrolled in the first term of intro-physics either graduated in a STEM field or did not.  This plot shows the fraction of those students who graduated in STEM after starting out in STEM.  These graduation fractions are shown for NU CLASP and NonCLASP classes on the left and for OU on the right.  The error bars are standard errors.  Again, several different ways of separating the students into demographic groups are shown.}
\label{fig:Grad}
\end{figure*}

For STEM graduation we use the same years that we have been using previously and include all students who enrolled in the first term of the appropriate series and who had a STEM major upon that enrollment. We then find the fraction of those students who eventually graduated and who had at least one graduating major in a STEM field. These STEM graduation rates are shown in Fig. \ref{fig:Grad} for the same demographic groups that we have been using previously.  We find that all demographic groups shown were at least as likely to graduate in STEM if they took a CLASP course than if they took a NonCLASP course with most groups being distinctly more likely to stay in STEM and graduate.  Given what we already knew about the changes in STEM graduation at OU (increased graduation rates but not enough to approach equitable) we were surprised that STEM graduation after a CLASP course at NU is consistent with our definition of equitable, at least among the demographic groups shown.  Even though students taking CLASP courses at OU graduated at higher rates, the distribution still seems far from equitable.   

Although the data in Fig. \ref{fig:Grad} may be used to compare graduation rates across demographics in order to judge whether a course is equitable, the absolute values of these graduation rates are harder to analyze because the university-wide graduation rates at OU and at NU were each slowly increasing with time over the times in our database.  This means that the increases that appear to happen when courses are changed to the CLASP structure may simply reflect a time-dependent increase in the overall graduate rates.  For NU we have graduation data on enough students across the university so that we can control for the time dependent effects but we don't have detailed data on all of the students enrolled at OU during these times.  We will limit ourselves to analyzing the effects of CLASP at NU.

To better understand the higher graduation rates under CLASP at NU we will construct a kind of difference-in-difference model (difference between CLASP and NonCLASP on top of possible differences across time) to allow us to separate a possible time dependent background from our physics student data and minimize the effects of the time differences inherent to the changeover to CLASP.  For this model we choose a control group from the students at NU who did not ever take either the NonCLASP or the CLASP course and who were enrolled at some time in the years of our study, 2010-2017.  We can use ethnicity and gender directly in our analyses so our control group just needs to have a large randomly chosen set of students with a variety of these characteristics.  However, the graduation probability depends rather strongly on other student characteristics such as academic year and whether the student began at NU as a Freshman or transferred from another school, so we would like a control group whose characteristics match those of our NonCLASP and CLASP courses. 
For these reasons, we identified a control group of 43,000 students chosen randomly but with weightings that gave us 15\% Freshmen, 33\% Sophomores, 26\% Juniors (12\% transfers and 14\% admitted earlier as Freshmen), and 26\% Seniors (14\% transfers and 12\% admitted as Freshmen).  Of that group we have ethnicity information for $N_{Control}=39,074$ control students.  We added this control group to our database of $N_{Physics}=2938$ NonCLASP and CLASP students who were undergraduates and for whom we have ethnicity data.  The control group includes both STEM majors and non-STEM majors so we will include both the STEM majors in the NU physics classes (68\% of the students) as well as the NonSTEM majors and the graduation rates here will not be STEM graduation but simply graduation in any major.

To account for changes in graduation rates over time, we use a hierarchical logistic regression model.  Specifically the students will be grouped according to the academic term of their entry into NU.  The hierarchical modeling fits each of these entry-term groups first and then assembles the resulting coefficients into overall final coefficients.  We use the student's first NU term to group them because each of these groups will be following the same sets of university rules related to graduation \footnote{We would get nearly the same results and so draw the same conclusions if we grouped them by enrollment term but entry term is more appropriate for graduation}.  Then we use the variable Control, = 1 for the control group and = 0 for the physics students, to distinguish the physics students in our database from the rest of the university.
We continue to use the variable CLASP, = 0 for students from NonCLASP classes and =1 for students from CLASP classes, to measure the difference between these types of class.  For each demographic group, we use this kind of hierarchical logistic regression method and fit the following model
\begin{equation}
\begin{split}
Log(GradOdds) & = b_0 + b_{Control}Control \\ & + b_{CLASP}CLASP
\end{split}
\label{eqn:GradRate}
\end{equation}
for the odds that a member of this group of students graduated.  For each group the fit produces the baseline overall graduation odds for NonCLASP students, the odds ratio comparing the control students to NonCLASP students, and the odds ratio comparing CLASP students to NonCLASP students.  With these odds and odds ratio estimates we can retrieve the average number of students per year who would have benefited from CLASP during these years. 

The coefficients resulting from the regression fit to Eq. \ref{eqn:GradRate} are given in Appendix \ref{sec:GradEqIndiv} in Table \ref{tab4a}.  We use those odds and odds ratios, along with Eq's \ref{eqn:Odds} and \ref{eqn:Ratio}, to find the graduation probabilities for each group in our model and show those probabilities in Table \ref{tab4} along with their standard errors.  Our main conclusions are seen in a comparison between CLASP and NonCLASP controlled graduation probabilities and in the estimated number of \textbf{additional} students graduating each year from NU who took the CLASP course.  Even after controlling for university-wide changes in graduation rates, each demographic group in Fig. \ref{fig:Grad} has a higher graduation probability if they took CLASP than that group had after a NonCLASP class.  The largest increase in controlled NU graduation rates came for female students from historically marginalized groups (examination of Fig. \ref{fig:Grad} suggests the same story would be true for OU).  Of course the result of these higher graduation rate estimates is that the extra number of CLASP students graduating is positive for each demographic group.  The data in Table \ref{tab4a} can also be used to find the ``graduation gap'' for historically marginalized students compared to NonHM. This gap is estimated to be 7.3\% $\pm$ 4\% for the control students, 6.0\% $\pm$ 2.3\% for the NonCLASP students, but only 1.0\% $\pm$ 3.4\% for the CLASP students.  We consider this evidence that the CLASP course structure is more equitable than the NonCLASP structure and nearly closes the previously existing graduation gap.   

\begin{table*}[htb]
\caption{\textbf{Class-type comparisons for graduation, for specific groups}.  The odds and odds ratios from fitting the logistic regression model, Eq. \ref{eqn:GradRate}, are shown in Appendix \ref{sec:GradEqIndiv}.  We use those odds and odds ratios, along with Eq's \ref{eqn:Odds} and \ref{eqn:Ratio}, to find the graduation probabilities for each group in our model.  We show these probabilities below with their standard errors.  The CLASP and NonCLASP graduation rates were then used to estimate the numbers of extra students graduating from CLASP each year that would not have graduated if they had taken the NonCLASP course.  The control group comprises over 92\% of \textit{N} for each demographic group.}
\label{tab4}
\begin{ruledtabular}
\begin{tabular}{c c | c c | c c | c c | c }
 & & \multicolumn{2}{c|}{\textbf{Control}} & \multicolumn{2}{c |}{\textbf{NonCLASP}} & \multicolumn{2}{c |}{\textbf{CLASP}} & \textbf{Extra CLASP} \\
\textbf{Group} & \textbf{\textit{N}} & \textbf{Grad.Prob.} & \textbf{Error} & \textbf{Grad.Prob.} & \textbf{Error} & \textbf{Grad.Prob.} & \textbf{Error} & \textbf{Grads/year (SE)}\\
 \hline
All &  46100 & 0.760 & 0.017 & 0.839 & 0.009 & 0.863 & 0.12 & 9.3 (6.9) \\
HM & 13222 & 0.712 & 0.035 & 0.791 & 0.020 & 0.853 & 0.029 & 5.9 (3.4) \\
NonHM &  28790 & 0.785 & 0.019 & 0.851 & 0.010 & 0.863 & 0.018 & 3.3 (5.7) \\
Female & 22809 & 0.784 & 0.023 & 0.861 & 0.012 & 0.894 & 0.018 & 6.1 (4.1) \\
Male & 23283 & 0.735 & 0.023 & 0.818 & 0.013 & 0.837 & 0.021 & 4.0 (5.3) \\
F\&HM & 7126 & 0.741 & 0.048 & 0.804 & 0.028 & 0.877 & 0.037 & 3.4 (2.2) \\
M\&HM & 6039 & 0.666 & 0.051 & 0.769 & 0.029 & 0.824 & 0.044 & 2.7 (2.6) \\
\end{tabular}
\end{ruledtabular}
\end{table*}

Again, these results are for all marginalized students who took CLASP at NU during these years and not just for STEM majors and, also, the rate we are considering is for graduation in any major.  In our entire dataset (which spans multiple years) there were 359 students from marginalized groups in CLASP classes so from a comparison of their CLASP graduation (in any major) rate of 0.853 with the same ethnicities’ NonCLASP graduation rate estimated to be 0.791 we see that 22 more students from marginalized groups graduated than one would predict if there had been no change to CLASP.  Interestingly, if we use the numbers in Fig. \ref{fig:Grad} and do the same calculation of how many ``extra'' \textbf{STEM majors} from marginalized groups graduated with a STEM major we find that the number is 21.  The fact that these numbers are almost the same suggests that the increase in the overall graduation rate that we found in our regression, for these particular students, is almost totally due to more CLASP students from marginalized groups choosing to stay in a STEM major and graduating with a STEM major.

\section{\label{sec:Discussion}Discussion}
This research is important for several different reasons. The first is that these kinds of studies are essential for progress to be made in higher education. While there is almost universal agreement across fields of education that active learning environments are better for learning and other important student outcomes, there is still extreme skepticism among many faculty, (in part because implementing active learning often requires that some content is removed from the course) which prevents widespread reform from the department level up. Demands for these active learning reforms also rarely come from the top down perhaps because of the large initial cost of investment \cite{brewe_costs_2018}. Therefore, one of the reasons research like this is useful is simply because it provides education research faculty an avenue to get peer-reviewed validation for their reform efforts to advocate for funding and convince their colleagues to continue teaching their courses using active learning. To be clear, research alone is not enough to change the culture around active learning, but it can be an essential step in the process.

Perhaps the reason that is most important to the authors is that it provides further evidence for the productive use of a course deficit model. As discussed in detail in our previous publication \cite{webb_attributing_2023}, there is a persistent myth that equity gaps exist because some groups of students come in with lesser experiences and that the course structure is either irrelevant or inviolable.  When one uses the student deficit model they generally ignore the possibility that the course controls these gaps.  In using the student deficit model, students are often referred to as ``not college ready'' or ``lacking preparation.''  Our view is that the data support a course deficit model explanation.  What a course deficit model does is reframe those equity gaps as a product of the course that is failing to leverage the assets of a diverse student population, rather than blaming the equity gaps on the students.  With this approach, the system must adapt to the changing population because the alternative is to force our students to adapt to an outdated inequitable system. Because we are able to reduce and in some cases eliminate equity gaps simply by changing the structure of a single introductory course, this is a strong indication that the university has much more power to eliminate equity gaps than they may realize.

This research provides additional evidence that introductory courses can potentially have a significant effect on student STEM retention, especially for marginalized students, and thus introductory courses are worth significantly investing in to boost long-term retention gains.

\section{\label{sec:Conclusions}Conclusions \& Future Directions}
Our results indicate that students are more likely to pass CLASP introductory physics courses than the traditional courses they replaced with 4.9\% more passing at Original University and 6.2\% more passing at Next University. They also indicate that students are less likely to drop CLASP courses with 59\% fewer students dropping at Original University and 81\% fewer dropping at next University. 

We also show that historically marginalized students who take CLASP are more likely to graduate with a STEM degree. Taking CLASP at Next university is correlated with a 6\% increase in graduation rate for HM students, and that is after controlling for the increase in graduation rates over time. This increase in historically marginalized student graduation rates effectively eliminates the HM student graduation gap for CLASP students. 

The CLASP students at both institutions perform as well as the students who took traditional physics on later coursework, despite the fact that many more students are passing CLASP than passed the traditional course. By using a model of Equity of Individuality \cite{Rodriguez2012} to frame our analysis, we claim that implementing CLASP is an equitable practice as it resulted in increased success for most of the demographics groups we studied and no significant decreases for any groups. In some cases, there is also a strong case to be made for meeting equity of parity, meaning HM students have similar outcomes to NonHM students.

There are also potentially other outside influences that may contribute to this change, but we have done our best to investigate those we know about. While skepticism is warranted around the claim that a single course can responsible for such a significant change in retention in the major, these findings are consistent with the findings of Hatfield, Brown \& Topaz \cite{Hatfield2022}. They found that a single grade lower than a C- reduces the probability of a student's graduation with a STEM degree by about 15\%.

This paper tells only a partial story regarding equity and success in CLASP. Pass rates for students in CLASP have changed in recent years at both institutions, indicating that some important elements of CLASP are tentative, and caution must be used before making claims that CLASP curriculum alone is responsible for improved success in physics and retention in STEM.  Nevertheless, the success of the CLASP curriculum for thousands of student over many years strongly supports the course deficit model of demographic grade gaps that we have previously argued for \cite{Paul2022,webb_attributing_2023} using different sets of course changes.

If we operate under the conclusion CLASP does improve student retention in STEM, we can currently make few claims about what aspects of the CLASP curriculum contribute to these outcomes. However, there are several potentially fruitful avenues to pursue to explain our results. For example, since CLASP students are actively engaged in model
construction and application instead of passively listening in lecture, they have more opportunities to successfully talk about science and use scientific tools. Moreover, these opportunities happen in a social
environment where students are communicating with instructors and peers. At Next University there is specific attention played to community building with activities throughout the semester that promote learning more about their classmates. This CLASP feature aligns with what research says is important for self-efficacy \cite{sawtelle_identifying_2012, Sawtelle2012}, identity \cite{carlone_understanding_2007} and belonging \cite{li_effects_2024} factors for persistence in STEM.

Future work will examine how these potential factors contribute to CLASP success. We will also examine how our results change with time, paying particular attention to grading practices in the CLASP and situate our findings in relevant race and identity theories \cite{carlone_understanding_2007}.

\section{\label{sec:acknowledgements}Acknowledgments} The authors would like to thank physics faculty at SJSU who reviewed an early draft of this work. This work is supported in part by NSF grant \# 1953760. 

\bibliography{ReplicationPaper}% Produces the bibliography via BibTeX.

\providecommand{\noopsort}[1]{}\providecommand{\singleletter}[1]{#1}%
\begin{thebibliography}{38}%
\makeatletter
\providecommand \@ifxundefined [1]{%
 \@ifx{#1\undefined}
}%
\providecommand \@ifnum [1]{%
 \ifnum #1\expandafter \@firstoftwo
 \else \expandafter \@secondoftwo
 \fi
}%
\providecommand \@ifx [1]{%
 \ifx #1\expandafter \@firstoftwo
 \else \expandafter \@secondoftwo
 \fi
}%
\providecommand \natexlab [1]{#1}%
\providecommand \enquote  [1]{``#1''}%
\providecommand \bibnamefont  [1]{#1}%
\providecommand \bibfnamefont [1]{#1}%
\providecommand \citenamefont [1]{#1}%
\providecommand \href@noop [0]{\@secondoftwo}%
\providecommand \href [0]{\begingroup \@sanitize@url \@href}%
\providecommand \@href[1]{\@@startlink{#1}\@@href}%
\providecommand \@@href[1]{\endgroup#1\@@endlink}%
\providecommand \@sanitize@url [0]{\catcode `\\12\catcode `\$12\catcode
  `\&12\catcode `\#12\catcode `\^12\catcode `\_12\catcode `\%12\relax}%
\providecommand \@@startlink[1]{}%
\providecommand \@@endlink[0]{}%
\providecommand \url  [0]{\begingroup\@sanitize@url \@url }%
\providecommand \@url [1]{\endgroup\@href {#1}{\urlprefix }}%
\providecommand \urlprefix  [0]{URL }%
\providecommand \Eprint [0]{\href }%
\providecommand \doibase [0]{https://doi.org/}%
\providecommand \selectlanguage [0]{\@gobble}%
\providecommand \bibinfo  [0]{\@secondoftwo}%
\providecommand \bibfield  [0]{\@secondoftwo}%
\providecommand \translation [1]{[#1]}%
\providecommand \BibitemOpen [0]{}%
\providecommand \bibitemStop [0]{}%
\providecommand \bibitemNoStop [0]{.\EOS\space}%
\providecommand \EOS [0]{\spacefactor3000\relax}%
\providecommand \BibitemShut  [1]{\csname bibitem#1\endcsname}%
\let\auto@bib@innerbib\@empty
%</preamble>
\bibitem [{\citenamefont {Schuette}(2023)}]{schuette_navigating_nodate}%
  \BibitemOpen
  \bibfield  {author} {\bibinfo {author} {\bibfnamefont {A.}~\bibnamefont
  {Schuette}},\ }\href@noop {} {\emph {\bibinfo {title} {Navigating the
  {Enrollment} {Cliff} in {Higher} {Education}}}},\ \bibinfo {type} {Tech.
  Rep.}\ (\bibinfo  {institution} {Trellis Company.},\ \bibinfo {year}
  {2023})\BibitemShut {NoStop}%
\bibitem [{\citenamefont {Copley}\ and\ \citenamefont
  {Douthett}(2020)}]{copley_enrollment_2020}%
  \BibitemOpen
  \bibfield  {author} {\bibinfo {author} {\bibfnamefont {P.}~\bibnamefont
  {Copley}}\ and\ \bibinfo {author} {\bibfnamefont {E.}~\bibnamefont
  {Douthett}},\ }\bibfield  {title} {\bibinfo {title} {The {Enrollment}
  {Cliff}, {Mega}-{Universities}, {COVID}-19, and the {Changing} {Landscape} of
  {U}.{S}. {Colleges}},\ }\href@noop {} {\bibfield  {journal} {\bibinfo
  {journal} {The CPA Journal}\ } (\bibinfo {year} {2020})}\BibitemShut
  {NoStop}%
\bibitem [{\citenamefont {Grawe}(2021)}]{grawe_how_2021}%
  \BibitemOpen
  \bibfield  {author} {\bibinfo {author} {\bibfnamefont {N.}~\bibnamefont
  {Grawe}},\ }\bibfield  {title} {\bibinfo {title} {How to survive the
  enrollment bust.},\ }\href@noop {} {\bibfield  {journal} {\bibinfo  {journal}
  {The Chronicle of Higher Education}\ }\textbf {\bibinfo {volume} {67}}
  (\bibinfo {year} {2021})}\BibitemShut {NoStop}%
\bibitem [{\citenamefont {Hatfield}\ \emph {et~al.}(2022)\citenamefont
  {Hatfield}, \citenamefont {Brown},\ and\ \citenamefont
  {Topaz}}]{Hatfield2022}%
  \BibitemOpen
  \bibfield  {author} {\bibinfo {author} {\bibfnamefont {N.}~\bibnamefont
  {Hatfield}}, \bibinfo {author} {\bibfnamefont {N.}~\bibnamefont {Brown}},\
  and\ \bibinfo {author} {\bibfnamefont {C.~M.}\ \bibnamefont {Topaz}},\
  }\bibfield  {title} {\bibinfo {title} {Do introductory courses
  disproportionately drive minoritized students out of stem pathways?},\ }\href
  {https://doi.org/10.1093/pnasnexus/pgac167} {\bibfield  {journal} {\bibinfo
  {journal} {PNAS Nexus}\ }\textbf {\bibinfo {volume} {1}},\ \bibinfo {pages}
  {1} (\bibinfo {year} {2022})}\BibitemShut {NoStop}%
\bibitem [{\citenamefont {Freeman}\ \emph {et~al.}(2014)\citenamefont
  {Freeman}, \citenamefont {Eddy}, \citenamefont {McDonough}, \citenamefont
  {Smith}, \citenamefont {Okoroafor}, \citenamefont {Jordt},\ and\
  \citenamefont {Wenderoth}}]{Freeman2014b}%
  \BibitemOpen
  \bibfield  {author} {\bibinfo {author} {\bibfnamefont {S.}~\bibnamefont
  {Freeman}}, \bibinfo {author} {\bibfnamefont {S.~L.}\ \bibnamefont {Eddy}},
  \bibinfo {author} {\bibfnamefont {M.}~\bibnamefont {McDonough}}, \bibinfo
  {author} {\bibfnamefont {M.~K.}\ \bibnamefont {Smith}}, \bibinfo {author}
  {\bibfnamefont {N.}~\bibnamefont {Okoroafor}}, \bibinfo {author}
  {\bibfnamefont {H.}~\bibnamefont {Jordt}},\ and\ \bibinfo {author}
  {\bibfnamefont {M.~P.}\ \bibnamefont {Wenderoth}},\ }\bibfield  {title}
  {\bibinfo {title} {Active learning increases student performance in science,
  engineering, and mathematics},\ }\href
  {https://doi.org/10.1073/pnas.1319030111} {\bibfield  {journal} {\bibinfo
  {journal} {Proceedings of the National Academy of Sciences}\ }\textbf
  {\bibinfo {volume} {111}},\ \bibinfo {pages} {8410} (\bibinfo {year}
  {2014})}\BibitemShut {NoStop}%
\bibitem [{\citenamefont {Kramer}\ \emph {et~al.}(2023)\citenamefont {Kramer},
  \citenamefont {Fuller}, \citenamefont {Watson}, \citenamefont {Castillo},
  \citenamefont {Oliva},\ and\ \citenamefont
  {Potvin}}]{kramer_establishing_2023}%
  \BibitemOpen
  \bibfield  {author} {\bibinfo {author} {\bibfnamefont {L.}~\bibnamefont
  {Kramer}}, \bibinfo {author} {\bibfnamefont {E.}~\bibnamefont {Fuller}},
  \bibinfo {author} {\bibfnamefont {C.}~\bibnamefont {Watson}}, \bibinfo
  {author} {\bibfnamefont {A.}~\bibnamefont {Castillo}}, \bibinfo {author}
  {\bibfnamefont {P.~D.}\ \bibnamefont {Oliva}},\ and\ \bibinfo {author}
  {\bibfnamefont {G.}~\bibnamefont {Potvin}},\ }\bibfield  {title} {\bibinfo
  {title} {Establishing a new standard of care for calculus using trials with
  randomized student allocation},\ }\href
  {https://doi.org/10.1126/science.ade9803} {\bibfield  {journal} {\bibinfo
  {journal} {Science}\ }\textbf {\bibinfo {volume} {381}},\ \bibinfo {pages}
  {995} (\bibinfo {year} {2023})},\ \bibinfo {note} {publisher: American
  Association for the Advancement of Science}\BibitemShut {NoStop}%
\bibitem [{\citenamefont {Docktor}\ and\ \citenamefont
  {Mestre}(2014)}]{Docktor2014}%
  \BibitemOpen
  \bibfield  {author} {\bibinfo {author} {\bibfnamefont {J.~L.}\ \bibnamefont
  {Docktor}}\ and\ \bibinfo {author} {\bibfnamefont {J.~P.}\ \bibnamefont
  {Mestre}},\ }\bibfield  {title} {\bibinfo {title} {Synthesis of
  discipline-based education research in physics},\ }\href
  {https://doi.org/10.1103/PhysRevSTPER.10.020119} {\bibfield  {journal}
  {\bibinfo  {journal} {Physical Review Special Topics - Physics Education
  Research}\ }\textbf {\bibinfo {volume} {10}},\ \bibinfo {pages} {1} (\bibinfo
  {year} {2014})}\BibitemShut {NoStop}%
\bibitem [{\citenamefont {Kanim}\ and\ \citenamefont {Cid}(2020)}]{Kanim2020}%
  \BibitemOpen
  \bibfield  {author} {\bibinfo {author} {\bibfnamefont {S.}~\bibnamefont
  {Kanim}}\ and\ \bibinfo {author} {\bibfnamefont {X.~C.}\ \bibnamefont
  {Cid}},\ }\bibfield  {title} {\bibinfo {title} {{Demographics of physics
  education research}},\ }\href
  {https://doi.org/10.1103/physrevphyseducres.16.020106} {\bibfield  {journal}
  {\bibinfo  {journal} {Phys. Rev. Phys. Educ. Res.}\ }\textbf {\bibinfo
  {volume} {16}},\ \bibinfo {pages} {020106} (\bibinfo {year}
  {2020})}\BibitemShut {NoStop}%
\bibitem [{\citenamefont {Potter}\ \emph {et~al.}(2014)\citenamefont {Potter},
  \citenamefont {Webb}, \citenamefont {Paul}, \citenamefont {West},
  \citenamefont {Bowen}, \citenamefont {Weiss}, \citenamefont {Coleman},\ and\
  \citenamefont {{De Leone}}}]{Potter2014Sixteen}%
  \BibitemOpen
  \bibfield  {author} {\bibinfo {author} {\bibfnamefont {W.}~\bibnamefont
  {Potter}}, \bibinfo {author} {\bibfnamefont {D.}~\bibnamefont {Webb}},
  \bibinfo {author} {\bibfnamefont {C.}~\bibnamefont {Paul}}, \bibinfo {author}
  {\bibfnamefont {E.}~\bibnamefont {West}}, \bibinfo {author} {\bibfnamefont
  {M.}~\bibnamefont {Bowen}}, \bibinfo {author} {\bibfnamefont
  {B.}~\bibnamefont {Weiss}}, \bibinfo {author} {\bibfnamefont
  {L.}~\bibnamefont {Coleman}},\ and\ \bibinfo {author} {\bibfnamefont
  {C.}~\bibnamefont {{De Leone}}},\ }\bibfield  {title} {\bibinfo {title}
  {{Sixteen years of collaborative learning through active sense-making in
  physics (CLASP) at UC Davis}},\ }\href {https://doi.org/10.1119/1.4857435}
  {\bibfield  {journal} {\bibinfo  {journal} {Am. J. Phys.}\ }\textbf {\bibinfo
  {volume} {82}},\ \bibinfo {pages} {153} (\bibinfo {year} {2014})}\BibitemShut
  {NoStop}%
\bibitem [{\citenamefont {Paul}\ \emph {et~al.}(2017)\citenamefont {Paul},
  \citenamefont {Webb}, \citenamefont {Chessey},\ and\ \citenamefont
  {Potter}}]{Paul2017}%
  \BibitemOpen
  \bibfield  {author} {\bibinfo {author} {\bibfnamefont {C.~A.}\ \bibnamefont
  {Paul}}, \bibinfo {author} {\bibfnamefont {D.~J.}\ \bibnamefont {Webb}},
  \bibinfo {author} {\bibfnamefont {M.~K.}\ \bibnamefont {Chessey}},\ and\
  \bibinfo {author} {\bibfnamefont {W.~H.}\ \bibnamefont {Potter}},\ }\bibfield
   {title} {\bibinfo {title} {Equity of success in {CLASP} courses at {UC}
  {Davis}},\ }\href {https://doi.org/10.1119/perc.2017.pr.068} {\bibfield
  {journal} {\bibinfo  {journal} {2017 Physics Education Research Conference
  Proceedings}\ ,\ \bibinfo {pages} {292}} (\bibinfo {year}
  {2017})}\BibitemShut {NoStop}%
\bibitem [{\citenamefont {Beichner}\ \emph {et~al.}(2007)\citenamefont
  {Beichner}, \citenamefont {Saul}, \citenamefont {Abbott}, \citenamefont
  {Morse}, \citenamefont {Allain}, \citenamefont {Bonham}, \citenamefont
  {Dancy},\ and\ \citenamefont {Risley}}]{beichner_student-centered_2007}%
  \BibitemOpen
  \bibfield  {author} {\bibinfo {author} {\bibfnamefont {R.~J.}\ \bibnamefont
  {Beichner}}, \bibinfo {author} {\bibfnamefont {J.~M.}\ \bibnamefont {Saul}},
  \bibinfo {author} {\bibfnamefont {D.~S.}\ \bibnamefont {Abbott}}, \bibinfo
  {author} {\bibfnamefont {J.~J.}\ \bibnamefont {Morse}}, \bibinfo {author}
  {\bibfnamefont {R.~J.}\ \bibnamefont {Allain}}, \bibinfo {author}
  {\bibfnamefont {S.~W.}\ \bibnamefont {Bonham}}, \bibinfo {author}
  {\bibfnamefont {M.~H.}\ \bibnamefont {Dancy}},\ and\ \bibinfo {author}
  {\bibfnamefont {J.~S.}\ \bibnamefont {Risley}},\ }\bibfield  {title}
  {\bibinfo {title} {The {Student}-{Centered} {Activities} for {Large}
  {Enrollment} {Undergraduate} {Programs} ({SCALE}-{UP}) {Project}},\
  }\bibfield  {journal} {\bibinfo  {journal} {Reviews in PER}\ }\textbf
  {\bibinfo {volume} {1}},\ \href
  {https://doi.org/http://www.per-central.org/document/ServeFile.cfm?ID=4517}
  {http://www.per-central.org/document/ServeFile.cfm?ID=4517} (\bibinfo {year}
  {2007})\BibitemShut {NoStop}%
\bibitem [{\citenamefont {Brewe}\ \emph {et~al.}(2010)\citenamefont {Brewe},
  \citenamefont {Sawtelle}, \citenamefont {Kramer}, \citenamefont {O’Brien},
  \citenamefont {Rodriguez},\ and\ \citenamefont
  {Pamelá}}]{brewe_toward_2010}%
  \BibitemOpen
  \bibfield  {author} {\bibinfo {author} {\bibfnamefont {E.}~\bibnamefont
  {Brewe}}, \bibinfo {author} {\bibfnamefont {V.}~\bibnamefont {Sawtelle}},
  \bibinfo {author} {\bibfnamefont {L.~H.}\ \bibnamefont {Kramer}}, \bibinfo
  {author} {\bibfnamefont {G.~E.}\ \bibnamefont {O’Brien}}, \bibinfo {author}
  {\bibfnamefont {I.}~\bibnamefont {Rodriguez}},\ and\ \bibinfo {author}
  {\bibfnamefont {P.}~\bibnamefont {Pamelá}},\ }\bibfield  {title} {\bibinfo
  {title} {Toward equity through participation in {Modeling} {Instruction} in
  introductory university physics},\ }\href
  {https://doi.org/10.1103/PhysRevSTPER.6.010106} {\bibfield  {journal}
  {\bibinfo  {journal} {Physical Review Special Topics - Physics Education
  Research}\ }\textbf {\bibinfo {volume} {6}},\ \bibinfo {pages} {010106}
  (\bibinfo {year} {2010})}\BibitemShut {NoStop}%
\bibitem [{\citenamefont {Redish}\ \emph {et~al.}(2014)\citenamefont {Redish},
  \citenamefont {Bauer}, \citenamefont {Carleton}, \citenamefont {Cooke},
  \citenamefont {Cooper}, \citenamefont {Crouch}, \citenamefont {Dreyfus},
  \citenamefont {Geller}, \citenamefont {Giannini}, \citenamefont {Gouvea},
  \citenamefont {Klymkowsky}, \citenamefont {Losert}, \citenamefont {Moore},
  \citenamefont {Presson}, \citenamefont {Sawtelle}, \citenamefont {Thompson},
  \citenamefont {Turpen},\ and\ \citenamefont
  {Zia}}]{redish_nexusphysics_2014}%
  \BibitemOpen
  \bibfield  {author} {\bibinfo {author} {\bibfnamefont {E.~F.}\ \bibnamefont
  {Redish}}, \bibinfo {author} {\bibfnamefont {C.}~\bibnamefont {Bauer}},
  \bibinfo {author} {\bibfnamefont {K.~L.}\ \bibnamefont {Carleton}}, \bibinfo
  {author} {\bibfnamefont {T.~J.}\ \bibnamefont {Cooke}}, \bibinfo {author}
  {\bibfnamefont {M.}~\bibnamefont {Cooper}}, \bibinfo {author} {\bibfnamefont
  {C.~H.}\ \bibnamefont {Crouch}}, \bibinfo {author} {\bibfnamefont {B.~W.}\
  \bibnamefont {Dreyfus}}, \bibinfo {author} {\bibfnamefont {B.~D.}\
  \bibnamefont {Geller}}, \bibinfo {author} {\bibfnamefont {J.}~\bibnamefont
  {Giannini}}, \bibinfo {author} {\bibfnamefont {J.~S.}\ \bibnamefont
  {Gouvea}}, \bibinfo {author} {\bibfnamefont {M.~W.}\ \bibnamefont
  {Klymkowsky}}, \bibinfo {author} {\bibfnamefont {W.}~\bibnamefont {Losert}},
  \bibinfo {author} {\bibfnamefont {K.}~\bibnamefont {Moore}}, \bibinfo
  {author} {\bibfnamefont {J.}~\bibnamefont {Presson}}, \bibinfo {author}
  {\bibfnamefont {V.}~\bibnamefont {Sawtelle}}, \bibinfo {author}
  {\bibfnamefont {K.~V.}\ \bibnamefont {Thompson}}, \bibinfo {author}
  {\bibfnamefont {C.}~\bibnamefont {Turpen}},\ and\ \bibinfo {author}
  {\bibfnamefont {R.~K.~P.}\ \bibnamefont {Zia}},\ }\bibfield  {title}
  {\bibinfo {title} {{NEXUS}/{Physics}: {An} interdisciplinary repurposing of
  physics for biologists},\ }\href {https://doi.org/10.1119/1.4870386}
  {\bibfield  {journal} {\bibinfo  {journal} {American Journal of Physics}\
  }\textbf {\bibinfo {volume} {82}},\ \bibinfo {pages} {368} (\bibinfo {year}
  {2014})}\BibitemShut {NoStop}%
\bibitem [{\citenamefont {Etkina}\ \emph {et~al.}(2019)\citenamefont {Etkina},
  \citenamefont {Brookes},\ and\ \citenamefont
  {Planinsic}}]{etkina_investigative_2019}%
  \BibitemOpen
  \bibfield  {author} {\bibinfo {author} {\bibfnamefont {E.}~\bibnamefont
  {Etkina}}, \bibinfo {author} {\bibfnamefont {D.~T.}\ \bibnamefont
  {Brookes}},\ and\ \bibinfo {author} {\bibfnamefont {G.}~\bibnamefont
  {Planinsic}},\ }\href {https://doi.org/10.1088/2053-2571/ab3ebd} {\emph
  {\bibinfo {title} {Investigative {Science} {Learning} {Environment}: {When}
  learning physics mirrors doing physics}}}\ (\bibinfo  {publisher} {Morgan \&
  Claypool Publishers},\ \bibinfo {year} {2019})\BibitemShut {NoStop}%
\bibitem [{\citenamefont {{Von Korff}}\ \emph {et~al.}(2016)\citenamefont {{Von
  Korff}}, \citenamefont {Archibeque}, \citenamefont {Gomez}, \citenamefont
  {Heckendorf}, \citenamefont {McKagan}, \citenamefont {Sayre}, \citenamefont
  {Schenk}, \citenamefont {Shepherd},\ and\ \citenamefont
  {Sorell}}]{VonKorff2016}%
  \BibitemOpen
  \bibfield  {author} {\bibinfo {author} {\bibfnamefont {J.}~\bibnamefont {{Von
  Korff}}}, \bibinfo {author} {\bibfnamefont {B.}~\bibnamefont {Archibeque}},
  \bibinfo {author} {\bibfnamefont {K.~A.}\ \bibnamefont {Gomez}}, \bibinfo
  {author} {\bibfnamefont {T.}~\bibnamefont {Heckendorf}}, \bibinfo {author}
  {\bibfnamefont {S.~B.}\ \bibnamefont {McKagan}}, \bibinfo {author}
  {\bibfnamefont {E.~C.}\ \bibnamefont {Sayre}}, \bibinfo {author}
  {\bibfnamefont {E.~W.}\ \bibnamefont {Schenk}}, \bibinfo {author}
  {\bibfnamefont {C.}~\bibnamefont {Shepherd}},\ and\ \bibinfo {author}
  {\bibfnamefont {L.}~\bibnamefont {Sorell}},\ }\bibfield  {title} {\bibinfo
  {title} {{Secondary analysis of teaching methods in introductory physics: A
  50 k-student study}},\ }\href {https://doi.org/10.1119/1.4964354} {\bibfield
  {journal} {\bibinfo  {journal} {Am. J. Phys.}\ }\textbf {\bibinfo {volume}
  {84}},\ \bibinfo {pages} {969} (\bibinfo {year} {2016})}\BibitemShut
  {NoStop}%
\bibitem [{\citenamefont {Etkina}\ \emph {et~al.}(1999)\citenamefont {Etkina},
  \citenamefont {Gibbons}, \citenamefont {Holton},\ and\ \citenamefont
  {Horton}}]{etkina_lessons_1999}%
  \BibitemOpen
  \bibfield  {author} {\bibinfo {author} {\bibfnamefont {E.}~\bibnamefont
  {Etkina}}, \bibinfo {author} {\bibfnamefont {K.}~\bibnamefont {Gibbons}},
  \bibinfo {author} {\bibfnamefont {B.~L.}\ \bibnamefont {Holton}},\ and\
  \bibinfo {author} {\bibfnamefont {G.~K.}\ \bibnamefont {Horton}},\ }\bibfield
   {title} {\bibinfo {title} {Lessons learned: {A} case study of an integrated
  way of teaching introductory physics to at-risk students at {Rutgers}
  {University}},\ }\href {https://doi.org/10.1119/1.19129} {\bibfield
  {journal} {\bibinfo  {journal} {American Journal of Physics}\ }\textbf
  {\bibinfo {volume} {67}},\ \bibinfo {pages} {810} (\bibinfo {year}
  {1999})}\BibitemShut {NoStop}%
\bibitem [{\citenamefont {Cooper}\ and\ \citenamefont
  {Brownell}(2016)}]{Cooper2016}%
  \BibitemOpen
  \bibfield  {author} {\bibinfo {author} {\bibfnamefont {K.~M.}\ \bibnamefont
  {Cooper}}\ and\ \bibinfo {author} {\bibfnamefont {S.~E.}\ \bibnamefont
  {Brownell}},\ }\bibfield  {title} {\bibinfo {title} {Coming {Out} in {Class}:
  {Challenges} and {Benefits} of {Active} {Learning} in a {Biology} {Classroom}
  for {LGBTQIA} {Students}},\ }\href {https://doi.org/10.1187/cbe.16-01-0074}
  {\bibfield  {journal} {\bibinfo  {journal} {CBE—Life Sciences Education}\
  }\textbf {\bibinfo {volume} {15}},\ \bibinfo {pages} {ar37} (\bibinfo {year}
  {2016})}\BibitemShut {NoStop}%
\bibitem [{\citenamefont {Shafer}\ \emph {et~al.}(2021)\citenamefont {Shafer},
  \citenamefont {Mahmood},\ and\ \citenamefont {Stelzer}}]{Shafer2021}%
  \BibitemOpen
  \bibfield  {author} {\bibinfo {author} {\bibfnamefont {D.}~\bibnamefont
  {Shafer}}, \bibinfo {author} {\bibfnamefont {M.~S.}\ \bibnamefont
  {Mahmood}},\ and\ \bibinfo {author} {\bibfnamefont {T.}~\bibnamefont
  {Stelzer}},\ }\bibfield  {title} {\bibinfo {title} {{Impact of broad
  categorization on statistical results: How underrepresented minority
  designation can mask the struggles of both Asian American and African
  American students}},\ }\href
  {https://doi.org/10.1103/PhysRevPhysEducRes.17.010113} {\bibfield  {journal}
  {\bibinfo  {journal} {Phys. Rev. Phys. Educ. Res.}\ }\textbf {\bibinfo
  {volume} {17}},\ \bibinfo {pages} {010113} (\bibinfo {year}
  {2021})}\BibitemShut {NoStop}%
\bibitem [{\citenamefont {Webb}\ and\ \citenamefont
  {Paul}(2023)}]{webb_attributing_2023}%
  \BibitemOpen
  \bibfield  {author} {\bibinfo {author} {\bibfnamefont {D.~J.}\ \bibnamefont
  {Webb}}\ and\ \bibinfo {author} {\bibfnamefont {C.~A.}\ \bibnamefont
  {Paul}},\ }\bibfield  {title} {\bibinfo {title} {Attributing equity gaps to
  course structure in introductory physics},\ }\href
  {https://doi.org/10.1103/PhysRevPhysEducRes.19.020126} {\bibfield  {journal}
  {\bibinfo  {journal} {Physical Review Physics Education Research}\ }\textbf
  {\bibinfo {volume} {19}},\ \bibinfo {pages} {020126} (\bibinfo {year}
  {2023})},\ \bibinfo {note} {publisher: American Physical Society}\BibitemShut
  {NoStop}%
\bibitem [{\citenamefont {Eddy}\ and\ \citenamefont
  {Hogan}(2014)}]{eddy_getting_2014}%
  \BibitemOpen
  \bibfield  {author} {\bibinfo {author} {\bibfnamefont {S.~L.}\ \bibnamefont
  {Eddy}}\ and\ \bibinfo {author} {\bibfnamefont {K.~A.}\ \bibnamefont
  {Hogan}},\ }\bibfield  {title} {\bibinfo {title} {Getting {Under} the {Hood}:
  {How} and for {Whom} {Does} {Increasing} {Course} {Structure} {Work}?},\
  }\href {https://doi.org/10.1187/cbe.14-03-0050} {\bibfield  {journal}
  {\bibinfo  {journal} {CBE—Life Sciences Education}\ }\textbf {\bibinfo
  {volume} {13}},\ \bibinfo {pages} {453} (\bibinfo {year} {2014})},\ \bibinfo
  {note} {publisher: American Society for Cell Biology (lse)}\BibitemShut
  {NoStop}%
\bibitem [{\citenamefont {Cotner}\ and\ \citenamefont
  {Ballen}(2017)}]{Cotner2017}%
  \BibitemOpen
  \bibfield  {author} {\bibinfo {author} {\bibfnamefont {S.}~\bibnamefont
  {Cotner}}\ and\ \bibinfo {author} {\bibfnamefont {C.~J.}\ \bibnamefont
  {Ballen}},\ }\bibfield  {title} {\bibinfo {title} {{Can mixed assessment
  methods make biology classes more equitable?}},\ }\href@noop {} {\bibfield
  {journal} {\bibinfo  {journal} {PLoS ONE}\ }\textbf {\bibinfo {volume}
  {12}},\ \bibinfo {pages} {e0189610} (\bibinfo {year} {2017})}\BibitemShut
  {NoStop}%
\bibitem [{\citenamefont {Valencia}(1997)}]{Valencia1997}%
  \BibitemOpen
  \bibfield  {author} {\bibinfo {author} {\bibfnamefont {R.~R.}\ \bibnamefont
  {Valencia}},\ }\href@noop {} {\emph {\bibinfo {title} {{The Evolution of
  Deficit Thinking: Educational Thought and Practice}}}}\ (\bibinfo
  {publisher} {The Falmer Press},\ \bibinfo {address} {London},\ \bibinfo
  {year} {1997})\BibitemShut {NoStop}%
\bibitem [{\citenamefont {Rodriguez}\ \emph {et~al.}(2012)\citenamefont
  {Rodriguez}, \citenamefont {Brewe}, \citenamefont {Sawtelle},\ and\
  \citenamefont {Kramer}}]{Rodriguez2012}%
  \BibitemOpen
  \bibfield  {author} {\bibinfo {author} {\bibfnamefont {I.}~\bibnamefont
  {Rodriguez}}, \bibinfo {author} {\bibfnamefont {E.}~\bibnamefont {Brewe}},
  \bibinfo {author} {\bibfnamefont {V.}~\bibnamefont {Sawtelle}},\ and\
  \bibinfo {author} {\bibfnamefont {L.~H.}\ \bibnamefont {Kramer}},\ }\bibfield
   {title} {\bibinfo {title} {{Impact of equity models and statistical measures
  on interpretations of educational reform}},\ }\href
  {https://doi.org/10.1103/physrevstper.8.020103} {\bibfield  {journal}
  {\bibinfo  {journal} {Phys. Rev. ST Phys. Educ. Res.}\ }\textbf {\bibinfo
  {volume} {8}},\ \bibinfo {pages} {020103+} (\bibinfo {year}
  {2012})}\BibitemShut {NoStop}%
\bibitem [{\citenamefont {Guti{\'{e}}rrez}(2012)}]{Gutierrez2012}%
  \BibitemOpen
  \bibfield  {author} {\bibinfo {author} {\bibfnamefont {R.}~\bibnamefont
  {Guti{\'{e}}rrez}},\ }\bibinfo {title} {Context matters: How should we
  conceptualize equity in mathematics education?},\ in\ \href
  {https://doi.org/10.1007/978-94-007-2813-4_2} {\emph {\bibinfo {booktitle}
  {Equity in Discourse for Mathematics Education}}},\ \bibinfo {series and
  number} {Mathematics Education Library},\ \bibinfo {editor} {edited by\
  \bibinfo {editor} {\bibfnamefont {B.}~\bibnamefont {Herbel-Eisenmann}},
  \bibinfo {editor} {\bibfnamefont {J.}~\bibnamefont {Choppin}}, \bibinfo
  {editor} {\bibfnamefont {D.}~\bibnamefont {Wagner}},\ and\ \bibinfo {editor}
  {\bibfnamefont {D.}~\bibnamefont {Pimm}}}\ (\bibinfo  {publisher}
  {Springer},\ \bibinfo {address} {Germany},\ \bibinfo {year} {2012})\ pp.\
  \bibinfo {pages} {17--33}\BibitemShut {NoStop}%
\bibitem [{\citenamefont {Paul}\ and\ \citenamefont {Webb}(2022)}]{Paul2022}%
  \BibitemOpen
  \bibfield  {author} {\bibinfo {author} {\bibfnamefont {C.~A.}\ \bibnamefont
  {Paul}}\ and\ \bibinfo {author} {\bibfnamefont {D.~J.}\ \bibnamefont
  {Webb}},\ }\bibfield  {title} {\bibinfo {title} {Percent grade scale
  amplifies racial or ethnic inequities in introductory physics},\ }\href
  {https://doi.org/10.1103/PhysRevPhysEducRes.18.020103} {\bibfield  {journal}
  {\bibinfo  {journal} {Phys. Rev. Phys. Educ. Res.}\ }\textbf {\bibinfo
  {volume} {18}},\ \bibinfo {pages} {020103} (\bibinfo {year}
  {2022})}\BibitemShut {NoStop}%
\bibitem [{\citenamefont {Harrer}\ and\ \citenamefont
  {Paul}(2019)}]{Harrer2019a}%
  \BibitemOpen
  \bibfield  {author} {\bibinfo {author} {\bibfnamefont {B.}~\bibnamefont
  {Harrer}}\ and\ \bibinfo {author} {\bibfnamefont {C.}~\bibnamefont {Paul}},\
  }\bibfield  {title} {\bibinfo {title} {Modeling {Energy} {Dynamics} with the
  {Energy}-{Interaction} {Diagram}},\ }\href
  {https://doi.org/10.1119/1.5126824} {\bibfield  {journal} {\bibinfo
  {journal} {The Physics Teacher}\ }\textbf {\bibinfo {volume} {57}},\ \bibinfo
  {pages} {462} (\bibinfo {year} {2019})}\BibitemShut {NoStop}%
\bibitem [{\citenamefont {Kim}\ and\ \citenamefont
  {Pak}(2002)}]{Kim2002Students}%
  \BibitemOpen
  \bibfield  {author} {\bibinfo {author} {\bibfnamefont {E.}~\bibnamefont
  {Kim}}\ and\ \bibinfo {author} {\bibfnamefont {S.-J.}\ \bibnamefont {Pak}},\
  }\bibfield  {title} {\bibinfo {title} {Students do not overcome conceptual
  difficulties after solving 1000 traditional problems},\ }\href
  {https://doi.org/10.1119/1.1484151} {\bibfield  {journal} {\bibinfo
  {journal} {American Journal of Physics}\ }\textbf {\bibinfo {volume} {70}},\
  \bibinfo {pages} {759} (\bibinfo {year} {2002})},\ \bibinfo {note}
  {publisher: AAPT}\BibitemShut {NoStop}%
\bibitem [{\citenamefont {Crouch}\ and\ \citenamefont
  {Mazur}(2001)}]{crouch_peer_2001}%
  \BibitemOpen
  \bibfield  {author} {\bibinfo {author} {\bibfnamefont {C.~H.}\ \bibnamefont
  {Crouch}}\ and\ \bibinfo {author} {\bibfnamefont {E.}~\bibnamefont {Mazur}},\
  }\bibfield  {title} {\bibinfo {title} {Peer {Instruction}: {Ten} years of
  experience and results},\ }\href {https://doi.org/10.1119/1.1374249}
  {\bibfield  {journal} {\bibinfo  {journal} {American Journal of Physics}\
  }\textbf {\bibinfo {volume} {69}},\ \bibinfo {pages} {970} (\bibinfo {year}
  {2001})}\BibitemShut {NoStop}%
\bibitem [{\citenamefont {{Castillo, Wendy}}\ and\ \citenamefont {{Gilborn,
  David}}(2023)}]{castillo_wendy_how_2023}%
  \BibitemOpen
  \bibfield  {author} {\bibinfo {author} {\bibnamefont {{Castillo, Wendy}}}\
  and\ \bibinfo {author} {\bibnamefont {{Gilborn, David}}},\ }\bibfield
  {title} {\bibinfo {title} {How to “{QuantCrit}:” {Practices} and
  {Questions} for {Education} {Data} {Researchers} and {Users}},\ }\bibfield
  {journal} {\bibinfo  {journal} {EdWorkingPaper:}\ }\textbf {\bibinfo {volume}
  {22-546}},\ \href {https://doi.org/10.26300/V5KH-DD65} {10.26300/V5KH-DD65}
  (\bibinfo {year} {2023}),\ \bibinfo {note} {publisher:
  EdWorkingPapers.com}\BibitemShut {NoStop}%
\bibitem [{APS(2024)}]{APSdiversity}%
  \BibitemOpen
  \href@noop {} {\bibinfo {title} {{American Association of Physics website
  regarding Diversity in Physics}}},\ \bibinfo {howpublished}
  {\url{https://www.aps.org/learning-center/statistics/diversity}} (\bibinfo
  {year} {2024}),\ \bibinfo {note} {accessed: 2024-06-25}\BibitemShut {NoStop}%
\bibitem [{Note1()}]{Note1}%
  \BibitemOpen
  \bibinfo {note} {Within the Course Deficit model the cause of the inequity is
  course structure. Hence the conclusion that the course is
  inequitable}\BibitemShut {NoStop}%
\bibitem [{Note2()}]{Note2}%
  \BibitemOpen
  \bibinfo {note} {We would get nearly the same results and so draw the same
  conclusions if we grouped them by enrollment term but entry term is more
  appropriate for graduation}\BibitemShut {NoStop}%
\bibitem [{\citenamefont {Brewe}\ \emph {et~al.}(2018)\citenamefont {Brewe},
  \citenamefont {Dou},\ and\ \citenamefont {Shand}}]{brewe_costs_2018}%
  \BibitemOpen
  \bibfield  {author} {\bibinfo {author} {\bibfnamefont {E.}~\bibnamefont
  {Brewe}}, \bibinfo {author} {\bibfnamefont {R.}~\bibnamefont {Dou}},\ and\
  \bibinfo {author} {\bibfnamefont {R.}~\bibnamefont {Shand}},\ }\bibfield
  {title} {\bibinfo {title} {Costs of success: {Financial} implications of
  implementation of active learning in introductory physics courses for
  students and administrators},\ }\href
  {https://doi.org/10.1103/PhysRevPhysEducRes.14.010109} {\bibfield  {journal}
  {\bibinfo  {journal} {Physical Review Physics Education Research}\ }\textbf
  {\bibinfo {volume} {14}},\ \bibinfo {pages} {010109} (\bibinfo {year}
  {2018})}\BibitemShut {NoStop}%
\bibitem [{\citenamefont {Sawtelle}\ \emph
  {et~al.}(2012{\natexlab{a}})\citenamefont {Sawtelle}, \citenamefont {Brewe},
  \citenamefont {Goertzen},\ and\ \citenamefont
  {Kramer}}]{sawtelle_identifying_2012}%
  \BibitemOpen
  \bibfield  {author} {\bibinfo {author} {\bibfnamefont {V.}~\bibnamefont
  {Sawtelle}}, \bibinfo {author} {\bibfnamefont {E.}~\bibnamefont {Brewe}},
  \bibinfo {author} {\bibfnamefont {R.~M.}\ \bibnamefont {Goertzen}},\ and\
  \bibinfo {author} {\bibfnamefont {L.~H.}\ \bibnamefont {Kramer}},\ }\bibfield
   {title} {\bibinfo {title} {Identifying events that impact self-efficacy in
  physics learning},\ }\href {https://doi.org/10.1103/PhysRevSTPER.8.020111}
  {\bibfield  {journal} {\bibinfo  {journal} {Physical Review Special Topics -
  Physics Education Research}\ }\textbf {\bibinfo {volume} {8}},\ \bibinfo
  {pages} {020111} (\bibinfo {year} {2012}{\natexlab{a}})}\BibitemShut
  {NoStop}%
\bibitem [{\citenamefont {Sawtelle}\ \emph
  {et~al.}(2012{\natexlab{b}})\citenamefont {Sawtelle}, \citenamefont {Brewe},\
  and\ \citenamefont {Kramer}}]{Sawtelle2012}%
  \BibitemOpen
  \bibfield  {author} {\bibinfo {author} {\bibfnamefont {V.}~\bibnamefont
  {Sawtelle}}, \bibinfo {author} {\bibfnamefont {E.}~\bibnamefont {Brewe}},\
  and\ \bibinfo {author} {\bibfnamefont {L.~H.}\ \bibnamefont {Kramer}},\
  }\bibfield  {title} {\bibinfo {title} {Exploring the relationship between
  self-efficacy and retention in introductory physics},\ }\href
  {https://doi.org/10.1002/tea.21050} {\bibfield  {journal} {\bibinfo
  {journal} {Journal of Research in Science Teaching}\ }\textbf {\bibinfo
  {volume} {49}},\ \bibinfo {pages} {1096} (\bibinfo {year}
  {2012}{\natexlab{b}})},\ \bibinfo {note} {publisher:
  Wiley-Blackwell}\BibitemShut {NoStop}%
\bibitem [{\citenamefont {Carlone}\ and\ \citenamefont
  {Johnson}(2007)}]{carlone_understanding_2007}%
  \BibitemOpen
  \bibfield  {author} {\bibinfo {author} {\bibfnamefont {H.~B.}\ \bibnamefont
  {Carlone}}\ and\ \bibinfo {author} {\bibfnamefont {A.}~\bibnamefont
  {Johnson}},\ }\bibfield  {title} {\bibinfo {title} {Understanding the science
  experiences of successful women of color: {Science} identity as an analytic
  lens},\ }\href {https://doi.org/10.1002/tea.20237} {\bibfield  {journal}
  {\bibinfo  {journal} {Journal of Research in Science Teaching}\ }\textbf
  {\bibinfo {volume} {44}},\ \bibinfo {pages} {1187} (\bibinfo {year}
  {2007})}\BibitemShut {NoStop}%
\bibitem [{\citenamefont {Li}\ \emph {et~al.}(2024)\citenamefont {Li},
  \citenamefont {Bernardi},\ and\ \citenamefont
  {Burkholder}}]{li_effects_2024}%
  \BibitemOpen
  \bibfield  {author} {\bibinfo {author} {\bibfnamefont {Y.}~\bibnamefont
  {Li}}, \bibinfo {author} {\bibfnamefont {R.~C.}\ \bibnamefont {Bernardi}},\
  and\ \bibinfo {author} {\bibfnamefont {E.}~\bibnamefont {Burkholder}},\
  }\bibfield  {title} {\bibinfo {title} {The effects of active learning on
  students’ sense of belonging and academic performance in introductory
  physics courses},\ }\href {https://doi.org/10.1088/1361-6404/ad4fcd}
  {\bibfield  {journal} {\bibinfo  {journal} {European Journal of Physics}\
  }\textbf {\bibinfo {volume} {45}},\ \bibinfo {pages} {045705} (\bibinfo
  {year} {2024})}\BibitemShut {NoStop}%
\bibitem [{Note3()}]{Note3}%
  \BibitemOpen
  \bibinfo {note} {When the number of students is small then the statistics are
  useless and we also risk identifying a student.}\BibitemShut {Stop}%
\end{thebibliography}%

\appendix

\section{\label{sec:DropPassGradEquity}Quantifying Equity of Parity}

\subsection{Dropping the Class}

As discussed in Sec.\ref{DropsWithin} our view of the Equity of Parity model of equity is that it matches Guti{\'{e}}rrez's \cite{Gutierrez2012} that demographic equity means that a student’s success shouldn’t be predictable from their demographic characteristics. Examination of Fig. \ref{fig:Drops} suggests that dropping a CLASP course has less dependence on demographics than dropping a NonCLASP course.  In this section we quantify this issue.  Specifically, for each independent demographic group in Fig. \ref{fig:Drops}, we will use a $\chi^2$ test to decide whether membership in that group is related to dropping the course.  The results of these tests, for both course types and at each university, are shown in Table \ref{tab2}.  When we find evidence that a group membership variable might be independent of the drop fraction variable (i.e. when the P-value is greater than 0.05), we place the associated large P-value in bold.  In addition to the $\chi^2$ results we also include, as an effect size, an estimate of the number of students, $n$, in the listed groups who \textbf{would not have dropped} if their group had the same drop fraction as their peers.  Table \ref{tab2} shows that, for either university, the CLASP courses more consistently produced these large P-values, suggesting that, by this particular measure, they were more generally equitable than the NonCLASP courses that they replaced. These statistical results are consistent with our view of Fig. \ref{fig:Drops} both overall, in showing CLASP courses seem more equitable, and in the details, for instance showing that the NonCLASP courses led to some groups of marginalized students being much more likely to drop than their peers.

\begin{table*}[htb] \caption{\textbf{Group-peer dropping the class comparisons within specific class-types}.  Results of $\chi^2$ tests of whether membership in the particular demographic group is statistically related to dropping the course.  Each test has one degree of freedom and the total number of students in the test, $N$ is shown as well as $\chi^2$ and the P-value.  Our definition of equity is when membership in a demographic group is not related to dropping the course so those tests with large P-values ($P>0.05$ noted in the table by bold numbers) denote groups that dropped the course about as often as their peers.  Also shown are effect sizes: $n$ = estimated number of extra students per year dropping the course (as determined from the group-nongroup inequality in drop fractions).  The standard error in $n$ is in parentheses.  The results shown in the left half of the table are for the NonCLASP and CLASP courses at Next University (NU) and the right half for the courses at Original University (OU).}
\label{tab2}
\begin{ruledtabular}
\begin{tabular}{c | c c c c | c c c c | c c c c | c c c c}
\textbf{Group} & \multicolumn{8}{c |}{\textbf{NU}} & \multicolumn{8}{c}{\textbf{OU}} \\ 
 & \multicolumn{4}{c}{\textbf{NonCLASP}} & \multicolumn{4}{c |}{\textbf{CLASP}} & \multicolumn{4}{c}{\textbf{NonCLASP}} & \multicolumn{4}{c}{\textbf{CLASP}} \\
 &  $N$ & $\chi^2$ & P & $n$(SE) & $N$ & $\chi^2$ & P & $n$(SE) & $N$ & $\chi^2$ & P & $n$(SE) & $N$ & $\chi^2$ & P & $n$(SE) \\
 \hline
 HM & 1758 & 5.47 & 0.02 & 2.8(1.4) & 1185 & 1.04 & \textbf{0.31} & \textbf{0.6(0.7)} & 3162 & 9.83 & 0.002 & 4.5(1.9) & 4586 & 1.12 & \textbf{0.29} & \textbf{0.8(0.9)} \\
 Female & 1927 & 0.35 & \textbf{0.55} & \textbf{1.2(2.0)} & 1292 & 2.11 & \textbf{0.15} & \textbf{1.6(1.1)} & 3256 & 2.42 & \textbf{0.12} & \textbf{7.9(4.7)} & 4740 & 1.24 & \textbf{0.27} & \textbf{-3.6(2.8)} \\
 F\&HM & 1758 & 0.74 & \textbf{0.39} & \textbf{0.7(0.9)} & 1185 & 1.27 & \textbf{0.26} & \textbf{0.4(0.5)} & 3162 & 12.7 & $<10^{-3}$ & 4.2(1.7) & 4586 & 3.57 & \textbf{0.06} & \textbf{1.1(0.8)} \\
M\&HM & 1758 & 4.81 & 0.03 & 1.8(1.0) & 1185 & 0.04 & \textbf{0.84} & \textbf{0.1(0.4)} & 3162 & 0.17 & \textbf{0.68} & \textbf{0.4(0.9)} & 4586 & 0.66 & \textbf{0.42} & \textbf{-0.4(0.4)} \\
\end{tabular}
\end{ruledtabular}
\end{table*}

\subsection{Passing the Class}

Again we use the model of Equity of Parity, this time to examine the data displayed in Fig. \ref{fig:Pass} and, again, argue that the CLASP courses appear to be the most equitable.  We quantify this assertion in the same way we did for data on dropping the course.  We use $\chi^2$ tests with each independent demographic group compared to their peers (the rest of the students).  The results of these tests are shown in Table \ref{tab3}.  When the P-values associated with these tests are large (in bold in the Table) we argue that the pass rate of the group is relatively close to that of their peers and we compare these results between CLASP and NonCLASP within NU or within OU.  Clearly, for CLASP courses, each of these demographic groups has pass rates close enough to their peers to yield large P-value results.  In contrast, several of the NonCLASP courses have rather smaller p-value in these tests.  Again, we don't use these results to argue that CLASP is fully equitable but just to argue that it appears more equitable than NonCLASP courses.

\begin{table*}[htb]
\caption{\textbf{Group-peer passing the class comparisons within specific class-types}.  Results of $\chi^2$ tests of whether membership in the particular demographic group is statistically related to passing the course with a grade of C- or higher.  Each test has one degree of freedom and the total number of students in the test, $N$ is shown as well as $\chi^2$ and the P-value.  Those tests with large P-values ($P>0.05$ noted in the table by bold numbers) denote groups that passed the course about as often as their peers.  Also shown are effect sizes: $n$ = estimated number of extra students per year who would pass the course (as determined from the group-nongroup inequality in pass fractions) if they had had the pass rates of their peer groups.  Values of $n$ which are small (i.e. within estimated errors of their peers) are indicative of group-peer equity.  The standard error in $n$ is in parentheses.  The results shown in the left half of the table are for the NonCLASP and CLASP courses at Next University (NU) and the right half for the courses at Original University (OU).}
\label{tab3}
\begin{ruledtabular}
\begin{tabular}{c | c c c c | c c c c | c c c c | c c c c}
\textbf{Group} & \multicolumn{8}{c |}{\textbf{NU}} & \multicolumn{8}{c}{\textbf{OU}} \\ 
 & \multicolumn{4}{c}{\textbf{NonCLASP}} & \multicolumn{4}{c |}{\textbf{CLASP}} & \multicolumn{4}{c}{\textbf{NonCLASP}} & \multicolumn{4}{c}{\textbf{CLASP}} \\
 & N & $\chi^2$ & P & n(SE) & N & $\chi^2$ & P & n(SE) & N & $\chi^2$ & P & n(SE) & N & $\chi^2$ & P & n(SE) \\
 \hline
HM & 1665 & 37.6 & $<10^{-3}$ & 11(2.1) & 1173 & 3.42 & \textbf{0.06} & \textbf{3.0(1.7)} & 3034 & 26.1 & $<10^{-3}$ & 9.3(2.4) & 4510 & 9.12 & 0.003 & 2.3(1.0) \\
 Female & 1821 & 0.19 & \textbf{0.66} & \textbf{-1.6(2.9)}  & 1278 & 2.31 & \textbf{0.13} & -\textbf{4.2(2.7)} & 3125 & 1.77 & \textbf{0.18} & \textbf{7.8(5.8)}  & 4663 & 3.28 & \textbf{0.07} & \textbf{5.3(2.8)} \\
 F\&HM & 1665 & 20.2 & $<10^{-3}$ & 5.1(1.4) & 1173 & 0.76 & \textbf{0.38} & \textbf{0.9(1.1)}  & 3034 & 21.9 & $<10^{-3}$ & 6.9(2.0) & 4510 & 2.04 & \textbf{0.15} & \textbf{0.9(0.7)}  \\
M\&HM & 1665 & 12.7 & $<10^{-3}$ & 4.4(1.4)  & 1173 & 2.28 & \textbf{0.13} & \textbf{1.6(1.2)}  & 3034 & 4.41 & 0.04 & 2.2(1.3) & 4510 & 8.75 & 0.003 & 1.3(0.7) \\
\end{tabular}
\end{ruledtabular}
\end{table*}

\section{\label{sec:GradEqIndiv}Graduation regressions - Equity of Individuality}

The odds ratios that were used to find the graduation probabilities in Table \ref{tab4} and resulted from fitting the data to Eq. \ref{eqn:GradRate} are given in Table \ref{tab4a}.  Our focus is on the CLASP odds ratios derived from $b_{CLASP}$ because those are (Odds graduating after taking CLASP)/(Odds graduating after taking NonCLASP).  As expected all of these odds ratios are larger than one so even after controlling for possible time dependence in overall graduation rates we still find that all of the demographic groups that we are considering graduated more often than the same groups taking NonCLASP classes.  The same extra numbers of CLASP students per year graduating as in Table \ref{tab4} are also shown in Table \ref{tab4a}.

\begin{table*}[htb]
\caption{\textbf{Class-type comparisons for graduation, for specific groups}.  Shown are the odds (from $b_0$), odds ratios (from $b_{Control}$ and $b_{CLASP}$), errors, and associated P-values resulting from our fit of the logistic regression model, Eq. \ref{eqn:GradRate} and Eq. \ref{eqn:Ratio}, to our data.  The control group comprises over 92\% of \textit{N} for each demographic group.  One can use the graduation odds for NonCLASP students together with the odds ratios for either the control students or for the CLASP students to find the graduation odds (and then the graduation rates) for those groups, respectively.  The CLASP graduation rates were used to estimate the numbers of extra students graduating from CLASP each year that would not have graduated if they had taken the NonCLASP course.  The standard errors are given in parentheses.}
\label{tab4a}
\begin{ruledtabular}
\begin{tabular}{c c | c c | c c | c c | c }
 & & \multicolumn{2}{c|}{\textbf{\textit{b}$_0$}} & \multicolumn{2}{c |}{\textbf{\textit{b}$_{Control}$}} & \multicolumn{2}{c |}{\textbf{\textit{b}$_{CLASP}$}} & \textbf{Extra CLASP}\\
\textbf{Group} & \textbf{\textit{N}} & \textbf{\textit{NonCLASPOdds}} (\textbf{SE}) & \textbf{P-value} & \textbf{\textit{OddsRatio}} (\textbf{SE}) & \textbf{\textit{P}}-value & \textbf{\textit{OddsRatio}} (\textbf{SE}) & \textbf{\textit{P}}-value & \textbf{Grad/year} (\textbf{SE})\\
 \hline
All &  46100 & 5.23 (0.35) & $<10^{-3}$ & 0.60 (0.04) & $<10^{-3}$ & 1.20 (0.12) & 0.073 & 9.3 (6.9) \\
HM &  13222 & 3.79 (0.47) & $<10^{-3}$ & 0.65 (0.08) & $<10^{-3}$ & 1.53 (0.30) & 0.026 & 5.9 (3.4) \\
NonHM &  28790 & 5.73 (0.47) & $<10^{-3}$ & 0.64 (0.05) & $<10^{-3}$ & 1.10 (0.14) & 0.444 & 3.3 (5.7) \\
Female & 22809 & 6.20 (0.60) & $<10^{-3}$ & 0.59 (0.06) & $<10^{-3}$ & 1.35 (0.22) & 0.067 & 6.1 (4.1) \\
 Male & 23283 & 4.50 (0.39) & $<10^{-3}$ & 0.62 (0.05) & $<10^{-3}$ & 1.14 (0.15) & 0.314 & 4.0 (5.3) \\
 F\&HM & 7126 & 4.11 (0.73) & $<10^{-3}$ & 0.70 (0.12) & 0.039 & 1.73 (0.51) & 0.060 & 3.4 (2.2) \\
M\&HM & 6039 & 3.34 (0.54) & $<10^{-3}$ & 0.60 (0.10) & 0.002 & 1.40 (0.36) & 0.189 & 2.7 (2.6) \\
\end{tabular}
\end{ruledtabular}
\end{table*}

\section{\label{sec:AllEthnicities}Ethnicity Lists from our OU and NU datasets}

As pointed out in the main text, the institutions ask each student to identify their ethnic background but each institution limits the data that we have access to by recording only one result for the category of ethnicity (although at NU there is a category for two or more ethnicities that does not specify any particular ethnicity).  In addition, the two institutions gather ethnicities in slightly different ways.  In our analyses in the main text we have combined students from marginalized ethnicities into a single group called HM but we wanted to show as much of the ungrouped data as made statistical sense to show.  In this appendix we will give a list of ethnicities included in our study as well as the label used for that group and, for groups making up at least 1\% of the total \footnote{When the number of students is small then the statistics are useless and we also risk identifying a student.}, the numbers of students present in the dataset.

First, Table \ref{tab:tabA1} identifies the symbols we use for the ethnicities at OU that are most populated in our dataset and make up 89\% of that dataset.  About 7\% of the students' ethnicities are unknown and there are four additional ethnicities represented in this OU dataset whose populations are too small, in either the NonCLASP classes or the CLASP classes or both, to be statistically useful.

\begin{table}[htb]
\caption{The ethnicities as described in the data obtained from Original University}
\label{tab:tabA1}
\begin{tabular}{c c}
\textbf{Symbol} & \textbf{Ethnicity} \\ 
\hline
AF & African-American/Black\\
CH & Chinese-American/Chinese\\
EI & East Indian/Pakistani \\
FP & Filipino/Filipino-American \\
IA & Indigenous American/American Indian/ \\
& Native American \\
JA  & Japanese-American/Japanese 
\\
KO & Korean-American/Korean \\
LA & Latino/Other Spanish \\
MX & Mexican-American/Mexican/ \\
& Chicano\\
OA & Other Asian \\
WH & White/Caucasian \\
\end{tabular}
\end{table}

Next, Table \ref{tab:tabA2} identifies the the ethnicities at NU that are most populated in our dataset and make up 81\% of that NU dataset.  About 9\% of the students' ethnicities are unknown and about 10\% are members of the 45 additional listed ethnicities at NU (most including fewer than ten students) represented in this NU dataset.  At this point we should note that we have previously shown \cite{Paul2022} that at least one of these ethnicities, Asian, contains many groups which don't all seem to experience the same biases.  Nevertheless, this grouping is already done for us in the NU dataset.

\begin{table}[htb]
\caption{The ethnicities as described in the data obtained from Next University}
\label{tab:tabA2}
\begin{tabular}{c c}
\textbf{Symbol} & \textbf{Ethnicity} \\ 
\hline
Asian & Asian\\
Blackprf & Black/African-American\\
 & Preference\\
Filipino & Filipino \\
Hispa & Hispanic/Latino \\
White & White \\
\end{tabular}
\end{table}

\section{\label{sec:DropsAllEthnicities}Dropping the course: More complete Ethincity\&Gender Data}

\begin{figure} [htb]
\includegraphics[trim=3.5cm 3.6cm 5.5cm 3.9cm, clip=true,width=\linewidth]{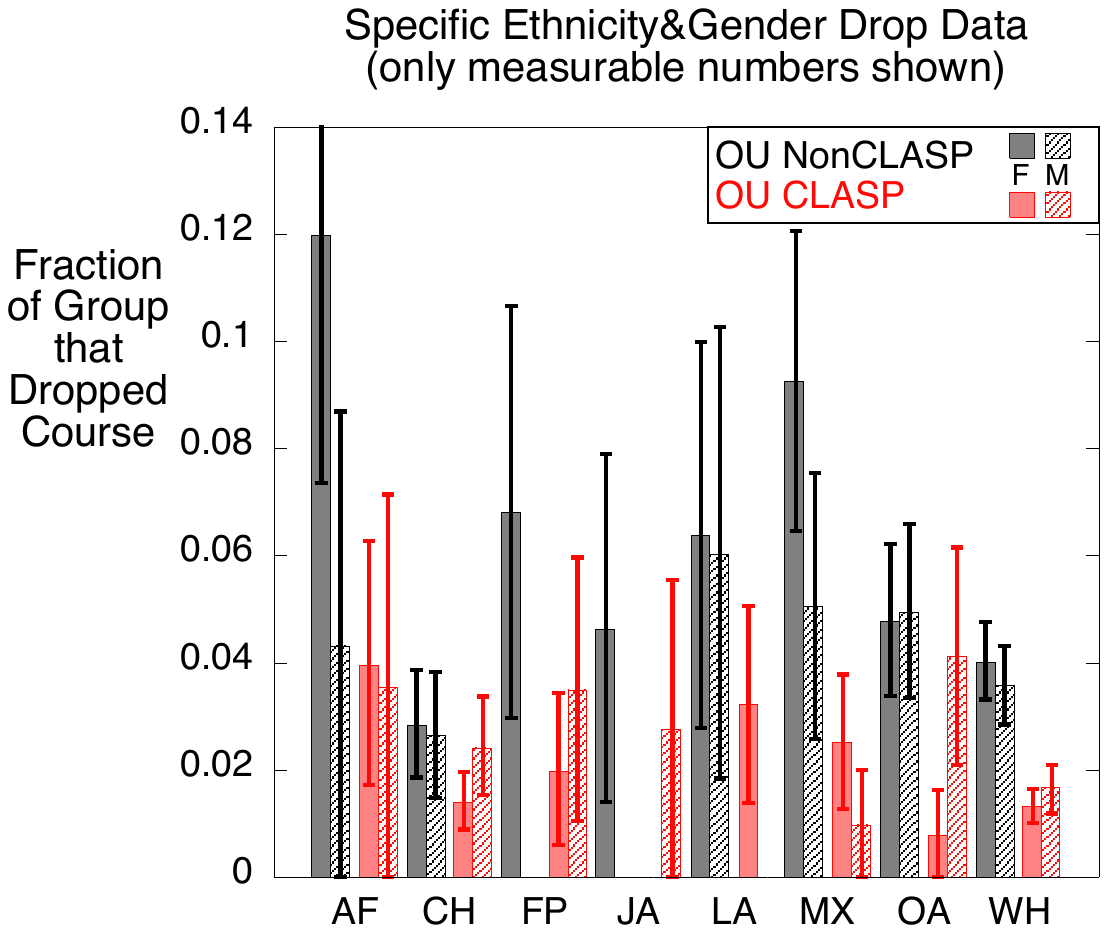}
\caption{Fraction of students dropping PhysicsA at OU separately showing i) the most populated ethnicities, ii) each ethnicity separated by gender, and iii) each Ethnicity\&Gender group drop fraction by class type (CLASP or NonCLASP).  The ethnicity symbols meanings are described in Table \ref{tab:tabA1}.  If a group from one of the ethnicities seems missing then that means that none of them dropped the course.}
\label{fig:EthGenOUDrops}
\end{figure}

A gender\&ethnicity and class-type breakdown of the fraction of students who drop the class is shown for OU in Fig. \ref{fig:EthGenOUDrops}.  We find that no Ethnicity\&Gender had significantly lower drops under NonCLASP courses.

\begin{figure} [htb]
\includegraphics[trim=3.4cm 4.7cm 5.5cm 3.6cm, clip=true,width=\linewidth]{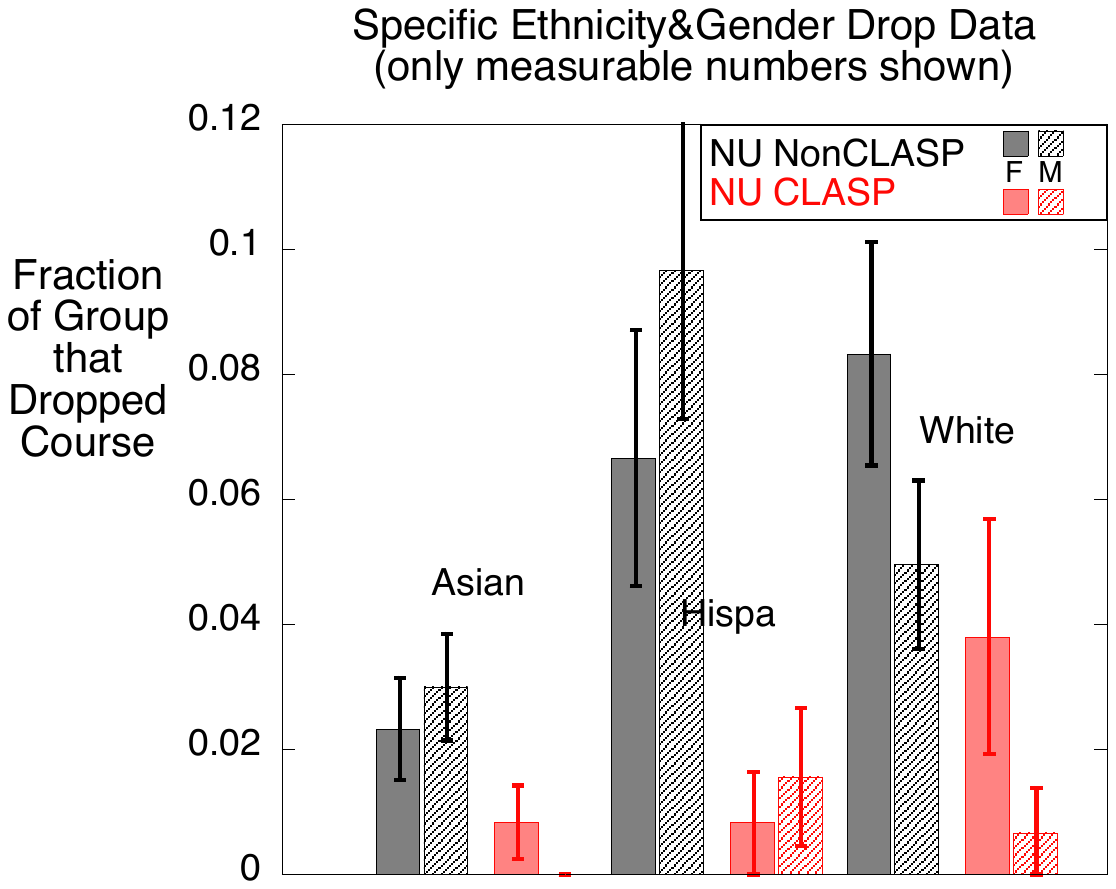}
\caption{Fraction of students dropping PhysicsA at NU separately showing i) the most populated ethnicities, ii) each ethnicity separated by gender, and iii) each Ethnicity\&Gender group drop fraction by class type (CLASP or NonCLASP).  The ethnicity symbols meanings are described in Table \ref{tab:tabA2}.  If a group from one of the ethnicities seems missing then that means that none of them dropped the course.}
\label{fig:EthGenNUDrops}
\end{figure}

A gender\&ethnicity and class-type breakdown of the fraction of students who drop the class is shown for NU in Fig. \ref{fig:EthGenNUDrops}.  We find that no Ethnicity\&Gender had significantly lower drops under NonCLASP courses.

\section{\label{sec:PassAllEthnicities}Passing the course: More complete Ethnicity\&Gender Data}

\begin{figure} [htbp]
\includegraphics[trim=3.5cm 3.6cm 5.5cm 3.9cm, clip=true,width=\linewidth]{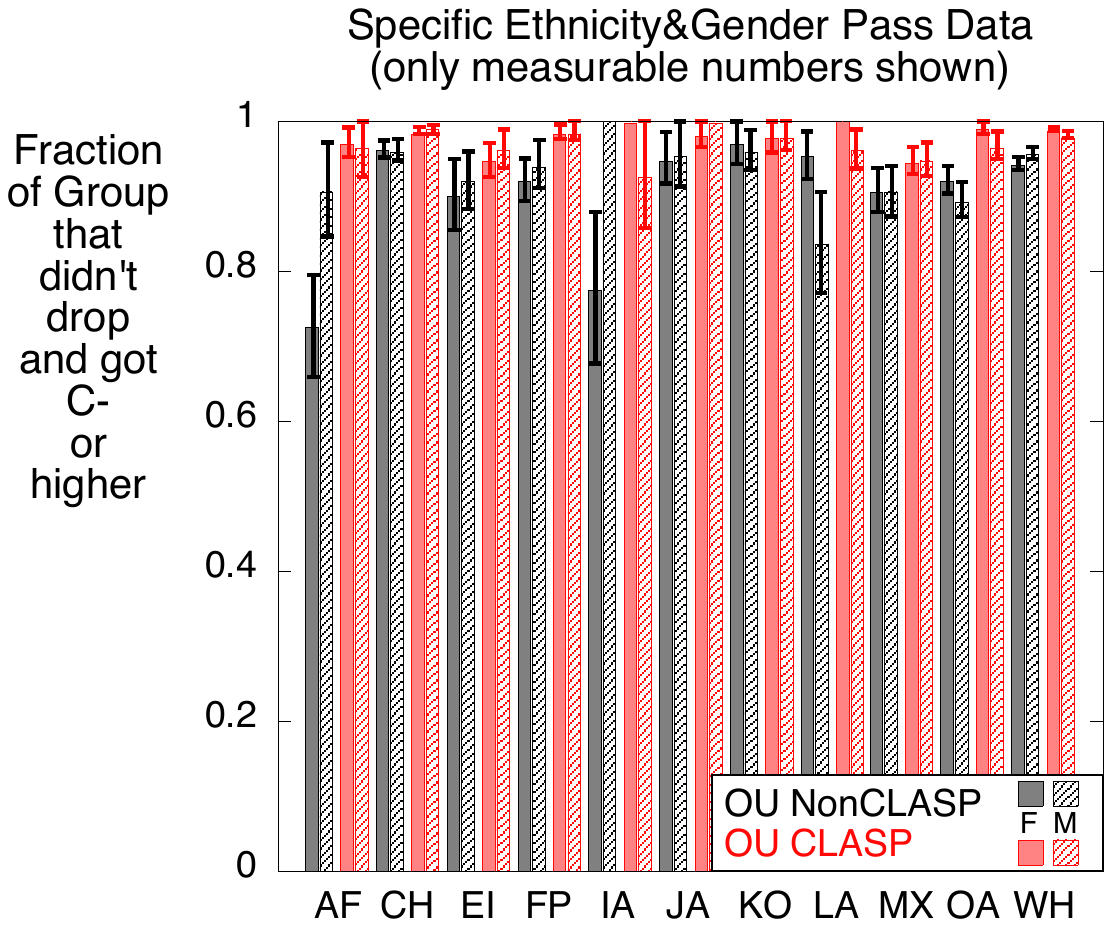}
\caption{Fraction of OU students who did not drop PhysicsA but did pass with a grade of C- or higher.  These are separately shown for i) the most populated ethnicities, ii) each ethnicity separated by gender, and iii) each Ethnicity\&Gender group drop fraction by class type (CLASP or NonCLASP).  The ethnicity symbols meanings are described in Table \ref{tab:tabA1}.}
\label{fig:EthGenOUPass}
\end{figure}

A gender\&ethnicity and class-type breakdown of the fraction of students at OU who did not drop PhysicsA and did receive a grade of C- or higher is shown in Fig. \ref{fig:EthGenOUPass}.  We find that no Ethnicity\&Gender had significantly higher pass rate under NonCLASP courses.

\begin{figure} [htb]
\includegraphics[trim=3.6cm 3.6cm 5.7cm 3.8cm, clip=true,width=\linewidth]{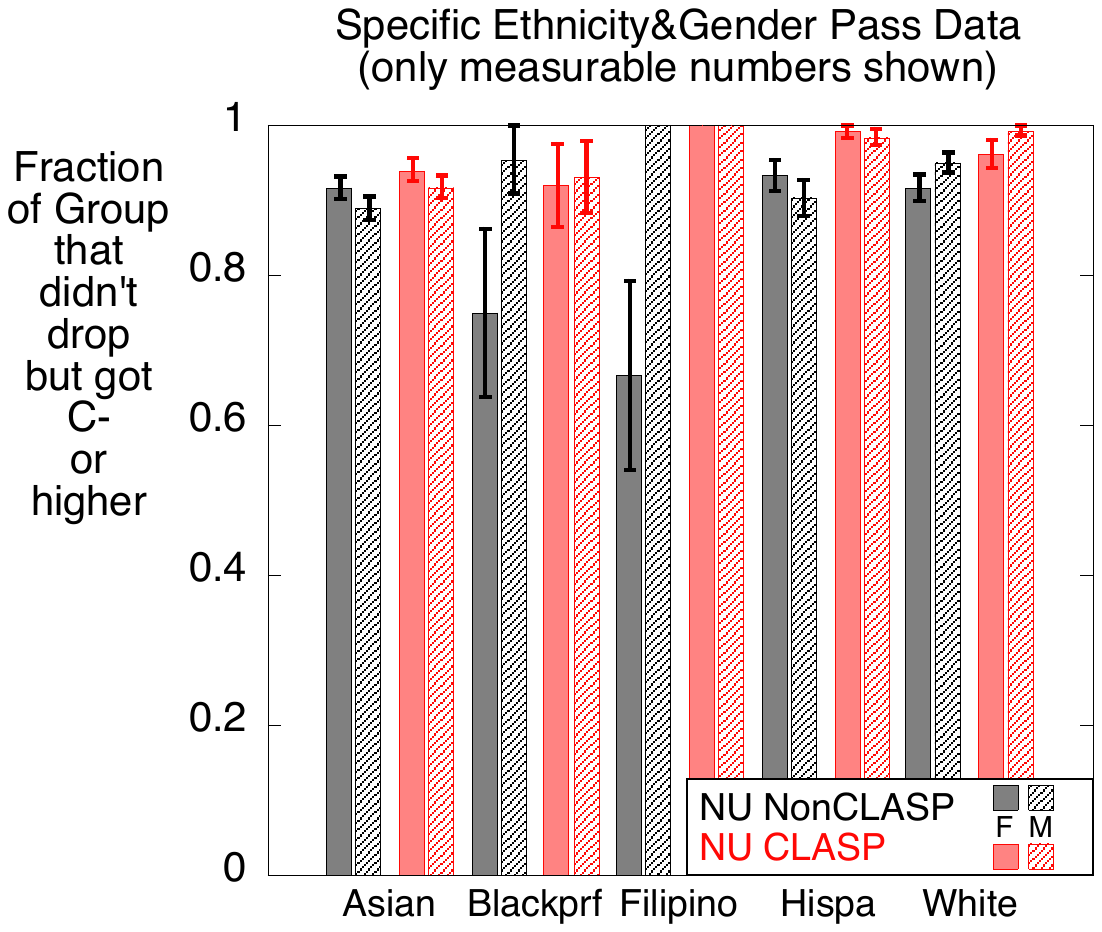}
\caption{Fraction of NU students who did not drop PhysicsA but did pass with a grade of C- or higher.  These are separately shown for i) the most populated ethnicities, ii) each ethnicity separated by gender, and iii) each Ethnicity\&Gender group drop fraction by class type (CLASP or NonCLASP).  The ethnicity symbols meanings are described in Table \ref{tab:tabA2}.}
\label{fig:EthGenNUPass}
\end{figure}

A gender\&ethnicity and class-type breakdown of the fraction of students at NU who did not drop PhysicsA and did receive a grade of C- or higher is shown in Fig. \ref{fig:EthGenNUPass}.  We find that no Ethnicity\&Gender had significantly higher pass rate under NonCLASP courses.

\section{\label{sec:GradAllEthnicities}STEM Graduation: More complete Ethincity\&Gender Data}

\begin{figure} [htb]
\includegraphics[trim=3.5cm 3.6cm 5.3cm 3.9cm, clip=true,width=\linewidth]{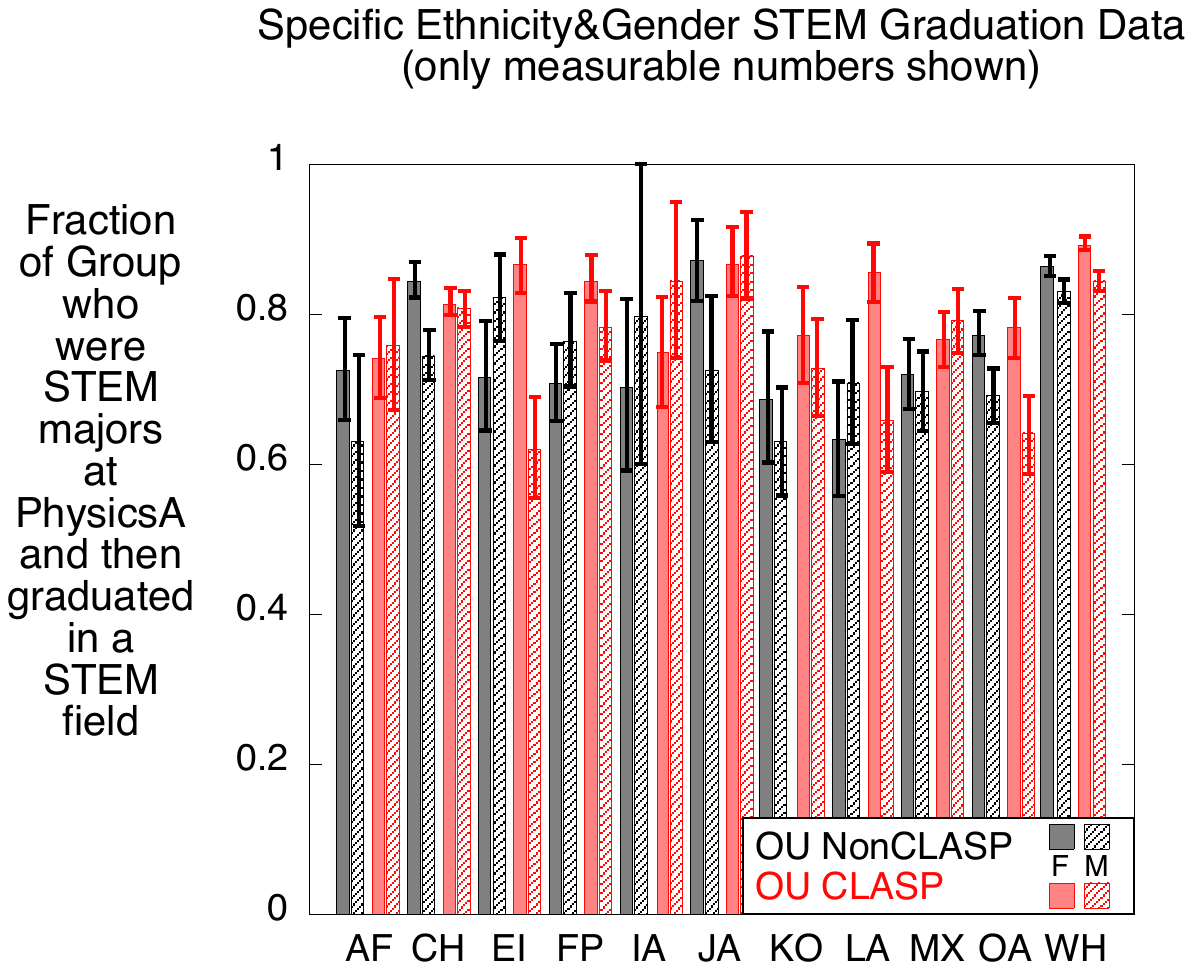}
\caption{Fraction of OU students who began PhysicsA as a STEM major and graduated with a STEM major.  These are separately shown for i) the most populated ethnicities, ii) each ethnicity separated by gender, and iii) each Ethnicity\&Gender group drop fraction by class type (CLASP or NonCLASP).  The ethnicity symbols meanings are described in Table \ref{tab:tabA1}.}
\label{fig:EthGenOUGrad}
\end{figure}

A gender\&ethnicity and class-type breakdown of the fraction of students at OU who began PhysicsA as a STEM major and graduated with a STEM major is shown in Fig. \ref{fig:EthGenOUGrad}.
Almost every Ethnicity\&Gender graduated in STEM at a higher rate under the OU CLASP courses and none passed at a significantly lower rate.

\begin{figure} [htb]
\includegraphics[trim=3.7cm 3.7cm 5.3cm 3.9cm, clip=true,width=\linewidth]{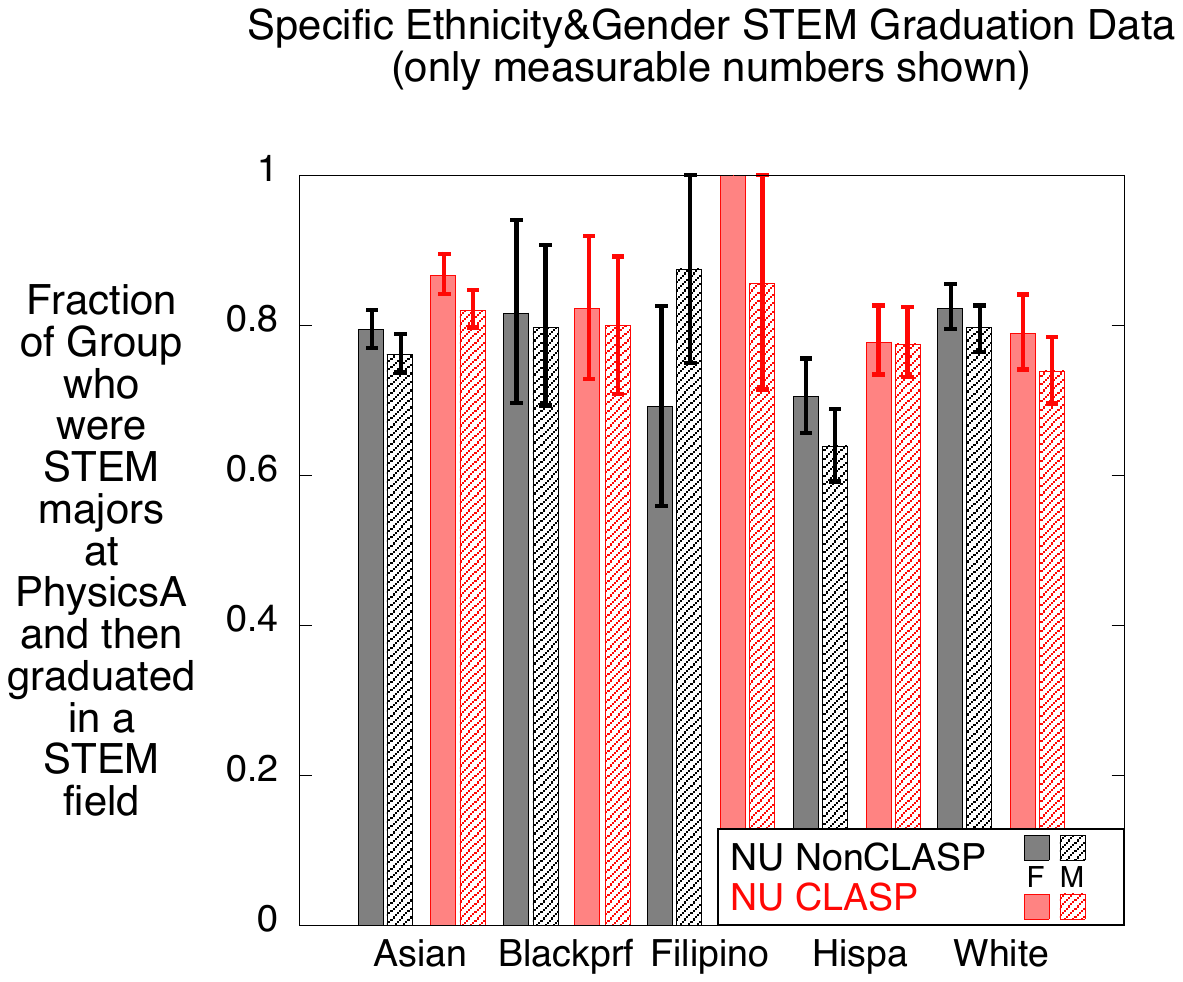}
\caption{Fraction of NU students who did not drop PhysicsA but did pass with a grade of C- or higher.  These are separately shown for i) the most populated ethnicities, ii) each ethnicity separated by gender, and iii) each Ethnicity\&Gender group drop fraction by class type (CLASP or NonCLASP).  The ethnicity symbols meanings are described in Table \ref{tab:tabA2}.}
\label{fig:EthGenNUGrad}
\end{figure}

A gender\&ethnicity and class-type breakdown of the fraction of students at NU who began PhysicsA as a STEM major and graduated with a STEM major is shown in Fig. \ref{fig:EthGenNUGrad}.  Almost every Ethnicity\&Gender graduated in STEM at a higher rate under the NU CLASP courses and none passed at a significantly lower rate.

%\bibliography{ReplicationPaper}% Produces the bibliography via BibTeX.

\end{document}